\documentclass{emulateapj}
\usepackage[utf8]{inputenc}
\usepackage[english]{babel}
\usepackage{fontenc}
\usepackage{graphicx}
\usepackage{color}

\newcommand{\galpak}{GalPaK$^{\rm 3D}$}

\newcommand{\FeII}{\hbox{{\rm Fe}{\sc \,ii}}}

\newcommand{\OII}{\hbox{{\rm O}{\sc \,ii}}}

\newcommand{\MgI}{\hbox{{\rm Mg}{\sc \,i}}}
\newcommand{\MgII}{\hbox{{\rm Mg}{\sc \,ii}}}

\newcommand{\HI}{\hbox{{\rm H}{\sc \,i}}}

\newcommand{\Ha}{\hbox{{\rm H}$\alpha$}}

\newcommand{\fdlam}{erg\,s$^{-1}$\,cm$^{-2}$\,\AA$^{-1}$}
\newcommand{\flux}{erg\,s$^{-1}$\,cm$^{-2}$}

\newcommand{\mpy}{\hbox{M$_{\odot}$\,yr$^{-1}$}}

\newcommand{\ma}{\hbox{$\lambda 2796$}}
\newcommand{\mb}{\hbox{$\lambda 2803$}}
\newcommand{\mc}{\hbox{$\lambda 2852$}}

\newcommand{\kms}{\hbox{${\rm km\,s}^{-1}$}}
\newcommand{\nn}{\nonumber}

\newcommand{\Vmax}{\hbox{$V_{\rm max}$}}
\newcommand{\Vout}{\hbox{$V_{\rm out}$}}
\newcommand{\thetam}{\hbox{$\theta_{\rm max}$}}

\begin{document}
\title{MusE GAs FLOw and Wind (MEGAFLOW) I: First MUSE results on background quasars~\altaffilmark{1}} 
\author{I.~{Schroetter\altaffilmark{2,4}},
        N.~{Bouch\'e\altaffilmark{3}}, 
        M.~{Wendt\altaffilmark{5,6}}, 
        T.~{Contini\altaffilmark{2,4}},
        H.~{Finley\altaffilmark{2,4}},
        R.~{Pell\'o\altaffilmark{2,4}},
        R.~{Bacon\altaffilmark{7}},
        S.~{Cantalupo\altaffilmark{9}},
        R.~A.~{Marino\altaffilmark{9}},
        J.~{Richard\altaffilmark{7}},
        S.~J.~{Lilly\altaffilmark{9}},
        J.~{Schaye\altaffilmark{8}},
        K.~{Soto\altaffilmark{9}},
        M.~{Steinmetz\altaffilmark{5}},
        L.~A.~{Straka\altaffilmark{8}},
        L.~{Wisotzki\altaffilmark{5}}}

\altaffiltext{1}{Based on observations made at the ESO telescopes under programs 094.A-0211(B) and 293.A-5038(A).}
\altaffiltext{2}{IRAP, Institut de Recherche en Astrophysique et Plan\'etologie, CNRS, 14, avenue Edouard Belin, F-31400 Toulouse, France}
\altaffiltext{3}{IRAP/CNRS, 9, avenue Colonel Roche, F-31400 Toulouse, France}
\altaffiltext{4}{University Paul Sabatier of Toulouse/ UPS-OMP/ IRAP, F-31400 Toulouse, France }
\altaffiltext{5}{AIP, Leibniz-Institut für Astrophysik Potsdam, An der Sternwarte 16, D-14482 Potsdam, Germany }
\altaffiltext{6}{Institut für Physik und Astronomie, Universität Potsdam, D-14476 Golm, Germany }
\altaffiltext{7}{Univ Lyon, Univ Lyon1, Ens de Lyon, CNRS, Centre de Recherche Astrophysique de Lyon UMR5574, F-69230, Saint-Genis-Laval, France }
\altaffiltext{8}{Leiden Observatory, Leiden University, PO Box 9513, 2300 RA Leiden, The Netherlands }
\altaffiltext{9}{ETH Zurich, Institute of Astronomy, Wolfgang-Pauli-Str. 27, 8093 Zurich, Switzerland }
\altaffiltext{10}{AIG, Institut für Astrophysik, Universität Göttingen, Friedrich-Hund-Platz 1, D-37077 Göttingen, Germany }

\begin{abstract}
The physical properties of galactic winds are one of the keys to understand galaxy formation and evolution.
These properties can be constrained thanks to  background quasar lines of sight (LOS) passing near star-forming galaxies (SFGs).
We present the first results of the MusE GAs FLOw and Wind (MEGAFLOW) survey obtained of 2 quasar fields  which have 8~\MgII\ 
absorbers of which 3 have rest-equivalent width greater than 0.8~\AA. 
With the new Multi Unit Spectroscopic Explorer (MUSE) spectrograph on the Very Large Telescope (VLT), 
we detect 6 (75$\%$)  \MgII\ host galaxy candidates withing a radius of $30\arcsec$ from the quasar LOS. 
Out of these 6 galaxy--quasar pairs, from geometrical arguments, one is likely probing galactic outflows, two are classified as ``ambiguous'', two are likely probing extended gaseous disks 
and one pair seems to be a merger. 
We focus on the wind$-$pair and constrain the outflow using a high resolution quasar spectra from Ultraviolet and Visual Echelle Spectrograph (UVES). 
Assuming the metal absorption to be due to gas flowing out of the detected galaxy through a cone along the minor axis, 
we find outflow velocities of the order of $\approx 150$~\kms\ (i.e. smaller than the escape velocity) with a loading factor, $\eta =\dot M_{\rm out}/$SFR, of $\approx$~0.7. 
We see evidence for an open conical flow, with a low-density inner core. 
In the future, MUSE will provide us with about 80 multiple galaxy$-$quasar pairs in two dozen fields. 

\end{abstract}


\section{Introduction}
\label{introduction}
\setcounter{footnote}{0}

In spite of the successes of the $\Lambda \rm CDM$ cosmological model \citep[i.e.][]{springel_05}, a major discrepancy remains between the predicted number density of 
dark matter halos and the observed number density of galaxies 
in the low-mass regime ($L<L_*$) \citep[i.e.][]{guo_10, papastergis_12, moster_10, moster_13, behroozi_13}.  
This behavior is usually explained by supernova(SN)-driven outflows \citep{dekel_86} which expel baryons from the galactic disk. 
Indeed, these galactic outflows are observed in almost every star-forming galaxy (SFG) \citep[][for a review]{VeilleuxS_05a} and are likely to enrich the inter-galactic 
medium \citep[e.g.][]{dekel_86, aguirre_01, oppenheimer_06}. 

The physical mechanisms for driving  galactic winds are complex and the cold gas
could be accelerated by  thermal energy injection \citep{springel_03}, 
by momentum injection from radiation pressure \citep[e.g.][]{murray_05}, 
by  cosmic ray pressure \citep[e.g.][]{booth_13, salem_14} 
or by a combination of these mechanisms \citep[e.g.][]{hopkins_15} 
The wide range physical scales that describe SN explosions from Astronomical Unit (AU) to tens of kiloparsecs (kpc), are beyond 
the capabilities of cosmological simulations. 

Hence, in most of these simulations, outflows are usually implemented with sub-grid prescriptions
\citep[e.g.][]{schaye_10, oppenheimer_10,Vogelsberger_14a}. 
A popular sub-grid recipe is to let the loading factor $\eta$, i.e. the ratio between the outflow rate $\dot M_{\rm out}$
and the star-formation rate (SFR), be a function of galaxy (halo) mass or circular velocity $V_c$ \citep{oppenheimer_10}
such as $\eta\propto V_c^{-1}$ for momentum-driven winds and $\eta\propto V_c^{-2}$ for energy-driven winds.
An alternative way to implement the collective effect of SN explosions is the (stochastic) implementation
of thermal feedback, where galactic winds develop without imposing any input outflow velocity nor mass loading factor
such as in the EAGLE simulations \citep[e.g.][]{schaye_15}, 
 the FIRE simulations \citep{hopkins_14, muratov_15}, 
and  the multi-phase scheme of \citet{barai_15}. 

Given the high impact of SN feedback on galaxy formation and the 
wide range of mass loading factors used in numerical simulations \citep[see the compilations in][]{zahid_14, torrey_14,schroetter_15},
observational constraints are of  paramount importance.
Unfortunately, our knowledge on the loading factor or the mass outflow rate $\dot M_{\rm out}$
is incomplete despite of the many efforts  made in the past decades \citep[i.e.][]{heckman_96, heckman_00, martin_98, martin_99, rupke_05, rubin_10, martin_12}. 
Indeed, estimates of the ejected mass flux $\dot M_{\rm out}$ using standard galaxy absorption lines 
\citep[e.g.][]{heckman_90, heckman_00, pettini_02, martin_02, martin_05, martin_12, martin_13}
 are uncertain by orders of magnitude mainly due to the difficulty in constraining the location of the probed outflowing gas~\footnote{
Furthermore, outflow rates from these low-ionization metal lines also require uncertain ionization corrections \citep[e.g.][]{chisholm_16}.
}. 
Indeed, the gas responsible for the blue shifted absorption lines in galaxies could be 0.1, 1 or 10 kpc away from the host.
Some recent studies have made serious attempts at determining the scaling of outflow rates with galaxy properties
 by setting the absorbing gas at a fixed distance \citep{ heckman_15, chisholm_15, wood_15}.

Background quasars can give us the minimum distance of the gas
from the impact parameter $b$ and thereby potentially yield more accurate outflow rates
\citep[][]{bouche_12, kacprzak_14, schroetter_15, muzahid_15}. 
One difficulty is that it is rare for the LOS to a background quasar to pass near a star-forming galaxy. 
Hence, one needs to devise strategies to build large samples of galaxy-quasar pairs. 
Another difficulty is that background quasars can probe not only 
the circum-galactic medium but also the outer regions of gaseous disks and the gas near other, undetected galaxies.

In order to obtain large samples of galaxy-quasar pairs, one can select quasars around galaxies or galaxies around quasars with absorption systems. 
The former requires quasar follow-up observations, while the latter requires one to detect the associated galaxies. 
In the era of large quasar catalogs from Sloan Digital Sky Survey (SDSS), we favor the absorption selection technique combined with integral field unit (IFU) observations.
Indeed, from \MgII\ absorption$-$selected quasar spectra, because we know the host galaxy redshift without knowing its position, IFUs can detect galaxies at previously unknown impact parameters. 
This kind of instrument also allows us to determine geometrical and kinematic properties of galaxies in the same observation. 
So far, IFUs such as SINFONI allowed us to probe galaxies within 20 kpc from the quasar line of sight (at redshift around 1). 
With the new VLT/MUSE instrument \citep{bacon_06, bacon_09}, one can now detect galaxies further away ($\sim$250 kpc away at $z=1$) thanks to its field of view 
of $1 \times 1$ arcmin (compared to $8 \arcsec \times8 \arcsec$ for SINFONI). 
The large wavelength coverage of MUSE (4800\AA\ to 9300\AA) allows us to target quasar fields with multiple \MgII\ ($\lambda \lambda 2796, 2802$) absorption lines 
having redshifts from 0.4 to 1.4 for [\OII]($\lambda \lambda 3727, 3729$) identification. 
We complement the VLT/MUSE IFU observations (which have a resolution $R\sim 2000$ or 150 \kms)
 with VLT/UVES follow-up high-resolution spectra of the quasars in order to study
 the line-of-sight kinematics with the resolution ($<10$ \kms) necessary for obtaining accurate constraints on outflow properties.

In this paper, we present the first results on galactic outflows from our MUSE survey. 
In \S~\ref{MEGAFLOW} we present the survey, the MUSE$+$UVES  data and the data reduction. 
\S~\ref{section:results_all} describes the sample results while \S~\ref{wind} presents our wind model as well as individual galaxy properties. 
Conclusions are then discussed in \S~\ref{conclusion}.

We use the $\Lambda\rm CDM$ standard cosmological parameters: $H_0$=70~km~s$^{-1}$, $\Omega_{\Lambda}$=0.7 and $\Omega_{M}$=0.3.
\section{The MEGAFLOW survey}
\label{MEGAFLOW}

\subsection{Target selection strategy}
\label{target_selection}

Current samples of galaxy$-$quasar pairs for strong \MgII\ absorbers, as in \citet{bouche_12, schroetter_15, muzahid_15} and \citet{bouche_16},
are made of a dozen pairs. 
Here, we seek to increase the sample size by almost an order of magnitude
in order to allow for statistical analysis of the relation between the absorption properties (and ultimately wind properties such as outflow rates and loading factors) and the galaxy properties. 
Thanks to the multiplexing capabilities of MUSE, 
having a sample 80---100 pairs is now within reach using 20--25 quasar fields.

As in our previous surveys,  we first select background quasar spectra with \MgII\ \ma\ absorption lines. 
For our MusE GAs FLOw and Wind (MEGAFLOW) survey, our strategy consists in selecting multiple \MgII\ absorbers (three, four or five) in quasar spectra 
from the Zhu and M\'enard catalog\footnote{This catalog can be found at http://www.pha.jhu.edu/~gz323/Site/Download$\_$Absorber$\_$Catalog.html} \citep{zhu_13} 
based on the SDSS survey \citep{ross_12, alam_15}. 
These \MgII\ absorptions should have redshifts between $0.4$ and $1.4$ such that the [\OII] $\lambda \lambda 3727, 3729$ galaxy emission lines fall into the MUSE wavelength range (4800\AA\ to 9300\AA).

To restrict the impact parameter range, we constrain the rest equivalent width (REW) of these absorptions $W_r^{\ma}$ to $W_r^{\ma}>$0.5~\AA\ 
because of the well-known anti-correlation between impact parameter and $W_r^{\ma}$ \citep{lanzetta_90, steidel_95, chen_10, kacprzak_11b, bordoloi_11, werk_13, nielsen_13}. 
Also the largest $W_r^{\ma}$ tend to be associated with outflows \citep[e.g.][]{kacprzak_11b, lan_14}. 
We define a strong absorber an absorber with $W_r^{\ma}>0.3 - 0.5$ \AA\ as in \citet{nestor_05}.
This limit of 0.5 \AA\ corresponds to $b \lesssim$100 kpc. 
We also need to pay attention to where the galaxy emission lines will appear in the spectrum and try to avoid bright sky emission lines as much as possible. 

The MEGAFLOW survey will consist of 20--25 quasar fields and the MUSE observations started in September 2014.
In October 2014, we obtained UVES observations on the first two fields
(Table~\ref{table:muse-obs})\footnote{as Director Discretionary Time (DDT) program 293.A-5038(A)}.
In this paper, we present the first results on these two fields towards SDSSJ213748$+$0012 and SDSSJ215200$+$0625, which
 have 4 \MgII\ absorption systems each.


\subsection{Observations and data reduction}
\label{section:observations}

\subsubsection{MUSE observations}
\label{observations}
MUSE data were taken in September 2014  in visitor mode during the first   Guaranteed Time Observations (GTO) run 
(program ID 0.94A-0211).  
We first point the telescope towards a quasar and then we offset the first exposure by $\approx 4-5 \arcsec$ in Right Ascension (RA) and Declination (Dec). 
This first offset is made to avoid the quasar flux to fall in the same pixels than the first pointing. 
Each observation was composed of four exposures of $900$ seconds with a rotation of $90^\circ$ between every exposure as well as small dithering ($<1\arcsec$).
This observation strategy is used in order to minimize systematics. 
From each MUSE observation, we obtain a combined cube of 317 $\times$ 316 spatial pixels (spaxels). 
Each spaxel has $\sim3680$ spectral pixels ranging from 4750~\AA\ to 9350~\AA. 
With a spectral sampling of 1.25~\AA$/$pixel, the average spectral resolution of the data is $\sim2.4$~\AA FWHM. 
The spatial resolution for the two quasar fields is $\sim0.8\arcsec$ FWHM with spatial sampling of 0.2$\arcsec/$pixel at 7000~\AA. 
The seeing constraint ($<0.9\arcsec$) is necessary if we want to derive galaxy parameters and detect them. 
Indeed, galaxies at redshift $\sim1$ can be small in size ($<1.2\arcsec$) and we need the seeing to be smaller than the galaxy to better derive its parameters. 

\begin{table*}
\centering
\caption{Summary of MUSE 094.A-0211(B) observations. \label{table:muse-obs}
}
\begin{tabular}{llcrc}
\hline
Field	 		          & $z_{\rm qso}$    &  PSF($\arcsec$)     &  T$_{\rm exp}$(s)  &  Date         \\    
(1)   		                  &	(2)	     & (3)	           &  (4)               &  (5)          \\    
\hline
J213748+0012                      &1.668             &0.8                  &3600                & 2014-09-23      \\  
J215200+0625                      &2.409             &0.7                  &7200                & 2014-09-24      \\  
\hline
\end{tabular}\\
\vspace*{0,1cm}
{
(1) Quasar name;
(2) Quasar emission redshift;
(3) FWHM of the seeing PSF (at $\approx$7000~\AA);
(4) Exposure time;
(5) Date of observations.
}\\   
\end{table*}

\subsubsection{MUSE data reduction}
\label{MUSE_reduction}
The data are reduced using version $1.0$ of the MUSE data reduction software (DRS) pipeline\footnote{A short description of the pipeline is given in \citet[][]{weilbacher_14}.}.
We process bias, flat field calibrations and arc lamp exposures taken during the night of the observations. 
Following calibration processing, raw science frames are bias subtracted and flat-fielded using master bias and master flat fields respectively.
The flat-fielding is renormalized in each slice to account for slight changes due to temperature variations
using a single flat field exposure taken hourly  before the science observation or when the instrument temperature changes
by more than 0.5$^{\circ}$ C.
An additional flat-field correction was performed using the twilight sky exposures taken at the beginning of each night  to correct for slight optical path differences between  sky and calibration unit. 
Geometrical calibration and astrometric solution are then applied.
The wavelength solution is obtained from the arc lamps and calibrated in air. Wavelengths are also corrected for the heliocentric velocity.
The flux calibration is obtained from a spectrophotometric star observed for each night. 

On each individual exposure, we use the default configuration of the DRS recipe and with the sky removal method turned off. 
This produces, for the $4$ individual exposures, a large table called the ``pixel-table''. 
For each individual exposure, star positions were registered in order to have accurate relative astrometry as shifts can occur 
between exposures due to the derotator wobble ($<0.3\arcsec$). 
The ``pixel-tables'' were then combined into a single data cube using the previously calculated offsets. 
The sky-subtraction was performed on this combined data cube with ZAP (Zurich Atmosphere Purge), an algorithm developed by \citet{soto_16b, soto_16a}. 
ZAP operates by first subtracting a baseline sky level, found by calculating the median per spectral plane
and  then uses principal component analysis  and determines the minimal number of eigenspectra that can reconstruct the residual emission features in the data cube.
Absolute astrometry is obtained by matching the positions of point sources in the data cube 
against the SDSS astrometry.

Finally, we cross$-$checked the flux calibration of these point sources against the SDSS magnitudes in the
$r$ and $i$ filter bands 
(the central wavelengths are $\lambda_r=6165 \rm\AA$ and $\lambda_i=7481 \rm\AA$ for $r$ and $i$ filters respectively)
 whose bandpass are within the MUSE wavelength coverage.
Using the $r$ and $i$ images obtained from the MUSE data cube convolved with the SDSS filters, 
we fitted a Moffat profile on each of the stars to calculate their total flux in 
each filter and then compare them with the SDSS ones. 
SDSS filters are design to be in AB magnitudes, but there are still corrections needed for some filters. 
Given that for the $r$ and $i$ filters, the AB to SDSS magnitudes correction is negligible,
we can correct fluxes into AB magnitudes directly using the following relation:
\begin{equation}
 AB = -2.5\log_{10}(f) - 5\log_{10}(<\lambda>) - 2.406 \label{eq:abmag}
\end{equation}
where $f$ is the flux in \fdlam\ and $<\lambda>$ the filter central wavelength in \AA. 

The comparison between MUSE and SDSS magnitudes is shown in Table~\ref{table:magnitudes}.
For both fields (J2137$+$0012 and J2152$+$0625), the agreement is around $1/10$th of a magnitude. 
In addition, another data reduction was performed using CubeFix and CubeSharp (Cantalupo, in prep) in order to show cleaner images 
of the fields in the Appendix (Fig~\ref{fig:J213748_field} and~\ref{fig:J215200_field}). 

\begin{table*}[]
\centering
\caption{Magnitude differences between MUSE and SDSS for J213748+0012 and J215200+0625 fields.}
\label{table:magnitudes}
\begin{tabular}{lccccccc}
\hline
Field		&object	&Instrument	&RA		& DEC				& mag$_r$	&mag$_i$			&Difference \\
(1)		&(2)	&(3)		&(4)		&(5)				&(6)		&(7)				&(8)			\\
\hline
J213748$+$0012	&QSO	&MUSE		&21:37:48.41	&$+$00:12:20.49				&18.33		&18.19					&$-0.13$\\
		&	&SDSS		&21:37:48.44	&$+$00:12:20.00				&18.20		&18.05					&\\
		&Star	&MUSE		&21:37:47.65	&$+$00:12:21.29				&19.71		&19.55					&$-0.09$\\
		&	&SDSS		&21:37:47.65	&$+$00:12:20.89				&19.61		&19.46					&\\
\hline
J215200$+$0625	&QSO	&MUSE		&21:52:00.05	&$+$06:25:17.26				&19.42		&19.44					&$-0.07$\\
		&	&SDSS		&21:52:00.03	&$+$06:25:16.36				&19.42		&19.30					&\\
		&Star	&MUSE		&21:51:59.84	&$+$06:25:05.48				&16.71		&16.47					&$-0.17$\\
		&	&SDSS		&21:51:59.83	&$+$06:25:04.72				&16.53		&16.29					&\\
\end{tabular}\\
{
(1) Field;
(2) Object type;
(3) Instrument (MUSE or SDSS);
(4) Right Ascension (RA);
(5) Declination (DEC); 
(6) Magnitude in $r$ filter (central wavelength $\lambda_r=6165 \rm\AA$);
(7) Magnitude in $i$ filter (central wavelength $\lambda_i=7481 \rm\AA$);
(8) Average difference SDSS$-$MUSE (mag).\\
}
\end{table*}

\subsubsection{UVES observation and reduction}
\label{section:uves}

The high resolution spectra for J213748+0012 and J215200+0625 were taken with 
UVES mounted on the 8.2m 
VLT at Paranal, Chile \citep{dekker_00}. 
These two fields were observed in DDT time under the program 293.A-5038(A). 
UVES is a cross-dispersed echelle spectrograph with two arms that are  
functionally identical: one covers the wavelengths in the  range 3000-5000 
\AA (Blue) and the other covers the range  4200-11000 \AA (Red). The 
details of the observational campaigns are presented in Table \ref{table:uves-obs}. 
The slit width of 1.2 arcsec and a CCD readout with 2x2 binning used for 
all the observations resulted in a spectral resolution power R $\approx 38000$ dispersed on pixels of $\sim$1.3 \kms. 
The settings were chosen in order to have a maximum of absorptions from host galaxies (from \FeII\ $\lambda2586$ to \MgI\ $\lambda2852$). 
The Common Pipeline Language (CPL version 6.3) of the UVES pipeline was 
used to bias correct and flat field the exposures and then to extract
the wavelength and flux calibrated spectra. After the standard reduction, the custom 
software  UVES popler\footnote{http$://$astronomy.swin.edu.au$/\sim$mmurphy$/$UVES$\_$popler$/$} 
(version 0.66) was used to combine the extracted echelle orders into 
single 1D spectra. The continuum was fitted with low-order polynomials.

\begin{table*}
\centering
\caption{Summary of UVES 293.A-5038(A) observations. \label{table:uves-obs}}
\begin{tabular}{llccc}
\hline
Target 		&setting $\lambda_c$ (nm)	&  T$_{\rm exp}$ (s)	&Date 		\\
\hline
J213748+0012	&390+580			&5970  			&2014-10-19	\\
J215200+0625	&390+580			&9015  			&2014-10-21,24 2014-11-18	\\
\hline
\end{tabular}

\vspace*{0,1cm}
\end{table*}

\section{MEGAFLOW sample first results}
\label{section:results_all}

\subsection{Galaxy detections}
\label{detection}

As we mentioned,  the two fields (SDSSJ213748$+$0012 and SDSSJ215200$+$0625)
 were selected to each have at least 3 absorbing systems with $W_r>0.5$~\AA\ (see Table~\ref{table:muse-det}). 

In each MUSE field, we search for [\OII] $\lambda \lambda 3727, 3729$ emission lines corresponding to the \MgII\ absorption redshifts seen in the quasar spectrum. 
However, the MUSE field of view of $1\arcmin \times 1\arcmin$ allows us to search for other companions in the fields, 
giving insight into the environment related to the host. 
We allow the potential host galaxies to have a redshift difference within a velocity interval of $\approx 1000$~\kms\ with respect to the absorber redshift ($z_{\rm gal} = z_{\rm abs} \pm 0.01$ for a $z\approx1$ galaxy). 
This velocity interval is set to prevent selection effects on surrounding gas velocities and thus not rejecting gas able to escape the gravitational well of the host galaxy in case of outflowing gas (more details on escape velocity in \S~\ref{outflow}). 
In the case where there are multiple galaxy candidates for a single \MgII\ line,
we select the galaxy with the smallest impact parameter from the quasar LOS. 
Table~\ref{table:muse-det} shows the detection rates for each field. 
For one of the undetected galaxies the expected emission line falls near a sky emission line at 7618 \AA\ 
(the $z\approx1.0437$ absorber in SDSSJ213748+0012) and the other line is too faint to be detected. 
For the reader interested in all of the galaxies detected in these MUSE data,  we provide  in the appendix a catalog with all the galaxies
for which a redshift could be determined.

\begin{table}
\centering
\caption{Summary of MUSE galaxy detection. \label{table:muse-det}}
\begin{tabular}{lccccc}
\hline
Field name 		&$z_{\rm absorber}$		& $W_{r}^{\lambda2796}$ 	&$N_{\rm det}$ 	&$b$	\\
(1)			&(2)				&(3)				&(4)			&(5)	\\
\hline
J213748+0012		&0.8063				&0.724$\pm$0.09			&1			&88	\\
			&1.0437				&0.767$\pm$0.08			&0\footnote{Affected by OH emission line at 7618\AA.}			&$\cdots$	\\
			&1.1890				&0.308$\pm$0.06			&1			&63	\\
			&1.2144				&1.144$\pm$0.06			&3			&87, 212, 246	\\
J215200+0625		&1.0534				&0.522$\pm$0.14			&2			&45, 189	\\
			&1.1761				&0.526$\pm$0.15			&0			&$\cdots$	\\
			&1.3190				&1.347$\pm$0.12			&1			&34	\\
			&1.4309				&1.152$\pm$0.11			&4			&62, 78, 184, 211	\\
\hline
\end{tabular}\\
{
(1) Quasar field name;
(2) \MgII\ absorption lines redshift;
(3) \MgII\ ($\lambda2796$) REW (\AA);
(4) Number of detected galaxies near absorber redshift;
(5) Impact parameter(s) of the detected galaxy(ies) (kpc);
}
\vspace*{0,1cm} 
\end{table}

We detect galaxies at redshifts of three of the four \MgII\ absorbers for the SDSSJ213748$+$0012 quasar field (see Table~\ref{table:muse-det}). 
For the \MgII\ absorber at $z= 0.8063$, we  find one [\OII] emission-line galaxy at a distance $b$ of 88~kpc. 
For the $z=1.1890$ \MgII\ absorber, we also find one galaxy at an impact parameter of 63~kpc. 
For the last $z=1.2144$ \MgII\ absorber, we find three [\OII] emitters, at  impact parameters of 87, 212 and 246~kpc. 
Given the large impact parameters of the latter two galaxies compared to the typical galaxy halo at these redshifts,
and given the large \MgII\ REW of 1 \AA, we assume the galaxy with the smallest impact parameter to be the host galaxy. 

For the SDSSJ215200$+$0625 field, we also detect galaxies at the redshifts of three out of the four \MgII\ absorbers (see Table~\ref{table:muse-det}). 
Two galaxies are identified for the first \MgII\ absorber at $z=1.0534$, at impact parameters of 45 and 189~kpc.
The host of the second absorber at $z=1.1761$ is not detected in spite of the wavelength for the expected [\OII] line being clear of OH lines.
The third \MgII\ absorption has a redshift of 1.3190 and has only one galaxy corresponding to that redshift at an impact parameter of 34~kpc. 
The last \MgII\ absorption is at $z=1.4309$ and we found 4 [\OII] emitters at that redshift,
which have impact parameters of 62, 78, 184 and 211~kpc (see Figure~\ref{fig:J2152G3}). 
This might be indicative of a group environment. 
Among, two have impact parameters very close to each other (62 and 78~kpc). 
We choose to assume that the closest galaxy (at 62~kpc) should be responsible for the \MgII\ 
absorption because it is the most massive and the brightest ($V_{\rm max} = 298$ \kms and 200 \kms, [\OII] fluxes being $5.05 \times 10^{-17}$\flux\ and $1.38 \times 10^{-17}$\flux , respectively). 

Using the propagated noise in the MUSE datacube, we can estimate flux (and surface brightness) limits on the expected [\OII] emission line for the non-detected host galaxies.
For the SDSSJ213748$+$0012 quasar field, at the first expected [\OII] wavelength ($\sim 6730 \AA$), with a noise of $2.3\times10^{-20}$ erg\,s$^{-1}$\,cm$^{-2}$\,\AA$^{-1}$ (1$\sigma$), 
we estimate a surface brightness limit of $1.43\times10^{-18}$ erg\,s$^{-1}$\,cm$^{-2}$\,arcsec$^{-2}$ (1$\sigma$) for emission line objects (assuming a $\rm FWHM = 2.48$ \AA). 
This corresponds to a flux limit of $1.04\times10^{-18}$ erg\,s$^{-1}$\,cm$^{-2}$ (1$\sigma$) for an unresolved emitter at 0.82$\arcsec$ seeing. 
The flux limit is $\sqrt{2}$ times for the [\OII] doublet (assuming a resolved doublet), or $1.47\times10^{-18}$ erg\,s$^{-1}$\,cm$^{-2}$ (1$\sigma$),
which corresponds to a SFR of 0.13~\mpy\ at $z=1$, typical of our sample. 
Surface brightness and flux limits are shown in Table~\ref{table:flux-limit}.

\subsection{SFR determination}

We use the $L_{\OII}$ ($\lambda \lambda 3727, 3729$) luminosity to estimate the SFR as follows.
We use the \citet{kennicutt_98} calibration, which assumes a \citet{salpeter_55} Initial Mass Function (IMF):
\begin{equation}
 SFR (\rm{M_{\odot}~yr^{-1}})=(1.4\pm0.4)\times 10^{-41}~ L([\OII])_o(\rm {erg}~s^{-1}) \label{eq:sfr}
\end{equation}
Where $L([\OII])_o$ is the [\OII] observed luminosity.
Using a \citet{chabrier_03} IMF and assuming a mean flux attenuation of $A_V=1$, which is typical for $z=1$ galaxies \citep[e.g.][]{charlot_02}, gives the same results (within 10\%) as Equation~\ref{eq:sfr}.

Equation 4 in \citet[][hereafter K04]{kewley_04} uses also a Salpeter IMF but makes no assumption of reddening. 
In their paper, they show that using the ``average'' attenuation correction of $0.3$ mag leads to underestimate the high SFR[\OII] ($>1$\mpy) and overestimate the low SFRs. 
They provide a way of deriving the E(B-V) (Eq.16 and 18 of K04) color excess which leads to a more accurate mean attenuation, assuming that $A_V=3.1\times E(B-V)$. 
We choose to use the following equations (Eq~\ref{eq:sfr_k04} and \ref{eq:sfr_k04_2} from K04) to derive our SFRs.
 
\begin{equation}
 SFR(\rm{M_{\odot}~yr^{-1}})=(6.58\pm1.65)\times 10^{-42}~ L([\OII])_i (\rm {erg}~s^{-1}) \label{eq:sfr_k04}
\end{equation}

\begin{equation}
 L([\OII])_i=3.11 \times 10^{-20}~ L([\OII])_o^{1.495} \label{eq:sfr_k04_2}
\end{equation}

\begin{table*}
\centering
\caption{Surface brightness and flux limits. \label{table:flux-limit} 
}
\begin{tabular}{lcccccc}
\hline
Quasar field 	&$z_{\rm absorber}$			&LSF	&  Noise 		&PSF	&Surface brightness limit 	&[\OII] flux limit	\\
(1) 		&(2)			&  (3)					&(4)	&(5)			&(6)	&(7)					\\
\hline
J213748+0012G1	&0.8063								&2.48	&$2.3\times10^{-20}$	&0.82 	&$1.43\times10^{-18}$		&$1.47\times10^{-18}$	\\
J213748+0012	&1.0437								&2.37	&$3.7\times10^{-20}$	&0.78 	&$2.19\times10^{-18}$		&$2.14\times10^{-18}$	\\
J213748+0012G2	&1.1890								&2.57	&$2.4\times10^{-20}$	&0.75 	&$1.54\times10^{-18}$		&$1.45\times10^{-18}$	\\
J213748+0012G3	&1.2144								&2.43	&$2.4\times10^{-20}$	&0.76 	&$1.45\times10^{-18}$		&$1.39\times10^{-18}$	\\
J215200+0625G1	&1.0534								&2.28	&$2.1\times10^{-20}$	&0.67 	&$1.19\times10^{-18}$		&$1.01\times10^{-18}$	\\
J215200+0625	&1.1761								&2.60	&$1.7\times10^{-20}$	&0.66 	&$1.10\times10^{-18}$		&$9.14\times10^{-19}$	\\
\textbf{J215200+0625G2}	&1.3190								&2.41	&$3.6\times10^{-20}$	&0.66 	&$2.17\times10^{-18}$		&$1.79\times10^{-18}$	\\
J215200+0625G3	&1.4309								&2.60	&$2.1\times10^{-20}$	&0.66 	&$1.36\times10^{-18}$		&$1.13\times10^{-18}$	\\
\hline
\end{tabular}\\
{
(1) Quasar field name;
(2) \MgII\ absorption line redshift;
(3) Line Spread Function FWHM (LSF) of the MUSE data (\AA);
(4) Data cube noise at the expected [\OII] wavelength (erg\,s$^{-1}$\,cm$^{-2}$\,\AA$^{-1}$) given at 1~$\sigma$;
(5) PSF of the data ($\arcsec$);
(6) Surface brightness limit (erg\,s$^{-1}$\,cm$^{-2}$\,arcsec$^{-2}$) given at 1~$\sigma$;
(7) [\OII] flux limit (erg\,s$^{-1}$\,cm$^{-2}$) given at 1~$\sigma$.\\
}
\end{table*}

\subsection{Galaxy morpho-kinematic properties}
\label{section:galaxy_properties}

Before classifying the galaxy$-$quasar pairs as favorable for gas outflows or inflows based on 
the azimuthal angle $\alpha$ of the apparent quasar location with respect to the galaxy major axis,
we need to determine the galaxy's major axis position angle (PA)\footnote{The position angle (PA) of a galaxy is the angle between the galaxy major axis and the celestial north.}. 

\begin{figure}[]
  \centering
  \includegraphics[width=4cm]{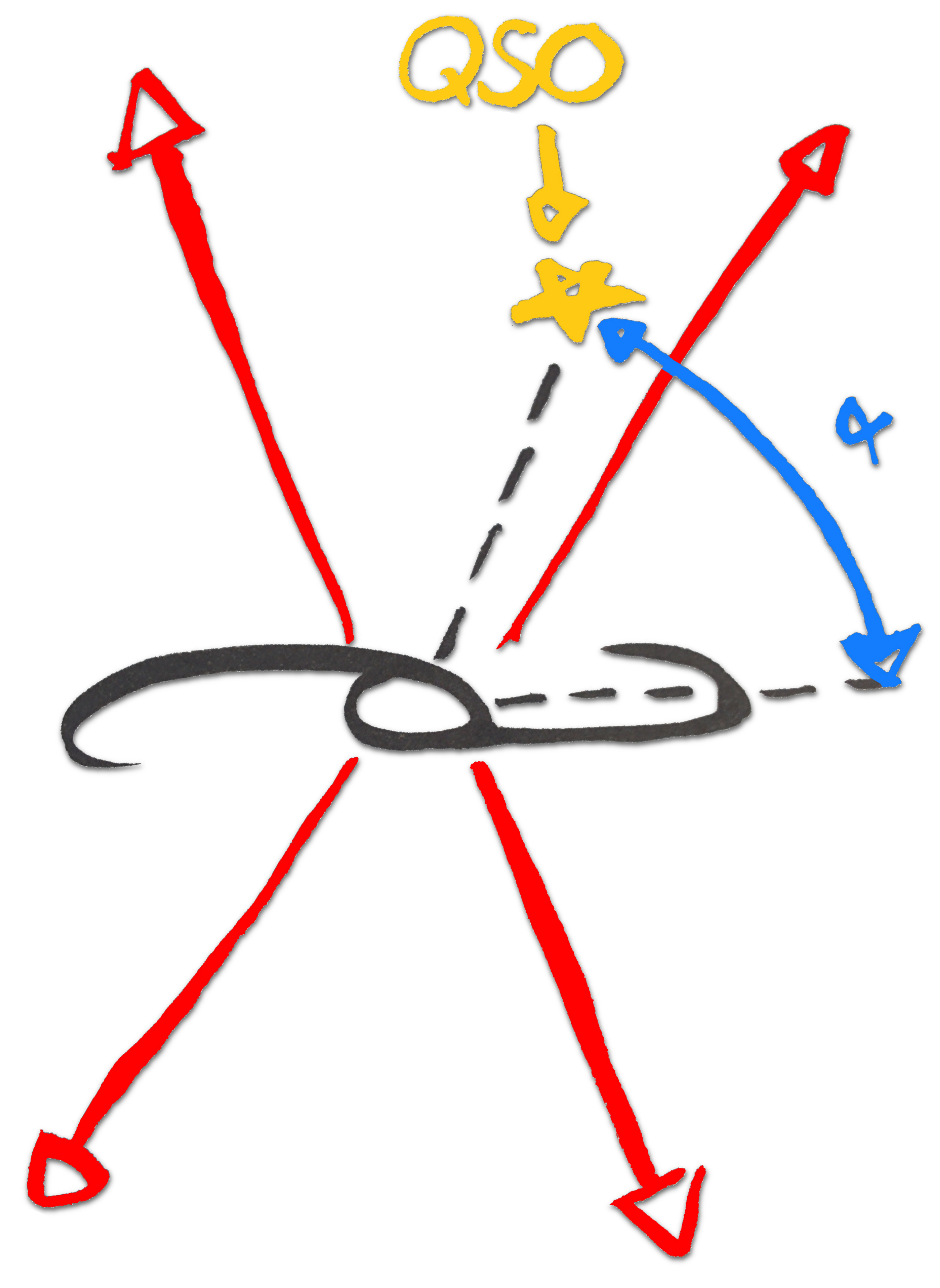}
  \caption{Scheme representing the azimuthal angle: 
  The galaxy is represented at the center in black, the red arrows represent the outflowing gas expelled from both side of the galaxy minor axis. 
  The azimuthal angle $\alpha$ is represented by the blue angle between the galaxy major axis and the quasar LOS (in yellow).
}
  \label{fig:alpha}
\end{figure}

We determine the PAs from  the morpho-kinematic properties of each galaxy using two approaches.
First, we used the 2D fitting tool Camel\footnote{The source code can be found at https://bitbucket.org/bepinat/camel.git} on the [\OII] emission lines 
to extract velocity and dispersion maps as in \citet{epinat_12} in order to establish whether the galaxy has a regular
velocity field compatible with a disk.  
Second, we use the \galpak\ algorithm \citep{bouche_15} to derive simultaneously the morphological and kinematic properties of these galaxies
using the continuum subtracted sub-cubes extracted around the [\OII] emission lines.
\galpak\  uses a disk parametric model with 10 free parameters and a Monte-Carlo Markov Chain (MCMC) algorithm with non-traditional sampling laws in order to efficiently probe the parameter space. Because the algorithm   uses  a 3-dimensional kernel
 to convolve the model with the spatial point-spread function (PSF or seeing) and the instrument line spread function (LSF), 
it returns the intrinsic (free of the PSF) galaxy properties (such as half-light radius, inclination, and maximum velocity).
Other parameters include the major-axis position angle, the galaxy flux, position, redshift and intrinsic velocity dispersion.
Results on the geometrical and kinematic properties of each galaxy are presented in Table~\ref{table:galpak}.

Figures~\ref{fig:J2137G1}$-$\ref{fig:J2152G3} 
show \galpak\ reconstructed models as well as Camel velocity maps for the 6 galaxies in the two fields. 
In Figure~\ref{fig:J2137G1} (SDSSJ213748+0012 field), the other emission sources are 
the quasar and a star's residual continuum.
In these figures, the left panel corresponds to a narrow band image of 30 pixels (37.5 \AA) around the galaxy's [\OII] emission lines. 
The background continuum has been subtracted so that we can only see the galaxy in emission.
In each of these Figures, we see the galaxy (inside the white rectangles) within 15$\arcsec$ of the quasar LOS (represented by a white cross). 
In the two right columns of these Figures, [\OII] integrated flux and velocity maps are shown. 
The top row corresponds to a 2$\times$2 (2 pixels FWHM) spatial Gaussian-smoothed flux map (left) and the Camel velocity map (right). 
The bottom row shows the \galpak\ model flux (left) and the PSF-deconvolved velocity (right) maps. 
We can see that in all cases, except in Figure~\ref{fig:J2137G2} for the dispersion-dominated SDSSJ213748+0012G2 galaxy,
 the model flux maps from \galpak\ is in a good agreement with the observed flux, and that \galpak\ and Camel velocity maps are consistent. 
Table~\ref{table:galpak} lists the resulting parameters for each galaxy.

\galpak\ results are reliable if the central galaxy pixel has, at minimum, a Signal to Noise Ratio (SNR) pixel$^{-1}$ of 3 \citep{bouche_15}. 
For each galaxy, we have SNR pixel$^{-1}$ of 11.0, 11.0, 4.5, 9.3, 4.2, 10.5 for SDSSJ213748+0012G1, G2, G3 and SDSSJ215200+0625G1, 
G2, G3 respectively. 
We checked that the parameters have converged for each galaxy as well as cross checked on raw data. 

\subsection{Classification and notes on the individual cases}

To put constraints on galactic outflows, we first need to select galaxy$-$quasar pairs suitable for wind studies (wind pairs). 
To do so, we measure the angle between the galaxy major axis and the apparent quasar location, which is
referred to as  the azimuthal angle $\alpha$ (see Figure~\ref{fig:alpha}).
Depending on this angle, the quasar LOS is likely to probe different phenomena around the galaxy.
If $55^\circ \leq \alpha \leq 90^\circ$, the quasar's position on the sky is roughly along the galaxy minor axis and is likely to cross the outflowing material of the 
galaxy\footnote{the Bordoloi papers have the definition of azimuthal angle reversed, i.e. their minor axis correspond to an $\alpha$ angle $<45^\circ$.} 
\citep[e.g.][]{bordoloi_11, bordoloi_14, kacprzak_12, kacprzak_14}.
If a pair has such an azimuthal angle, it will be classified as a wind-pair.
On the other hand, if the quasar is positioned along the galaxy major axis ($0^\circ \leq \alpha \leq 30^\circ$), the quasar LOS is likely to probe inflowing or circum-galactic gas.
With such configuration, we classify the pair as suitable for accretion studies (inflow pair). 
In between, ($35^\circ \leq \alpha \leq 55^\circ$), we cannot distinguish between these two extreme cases. 

\begin{figure}[]
  \centering
  \includegraphics[width=9cm]{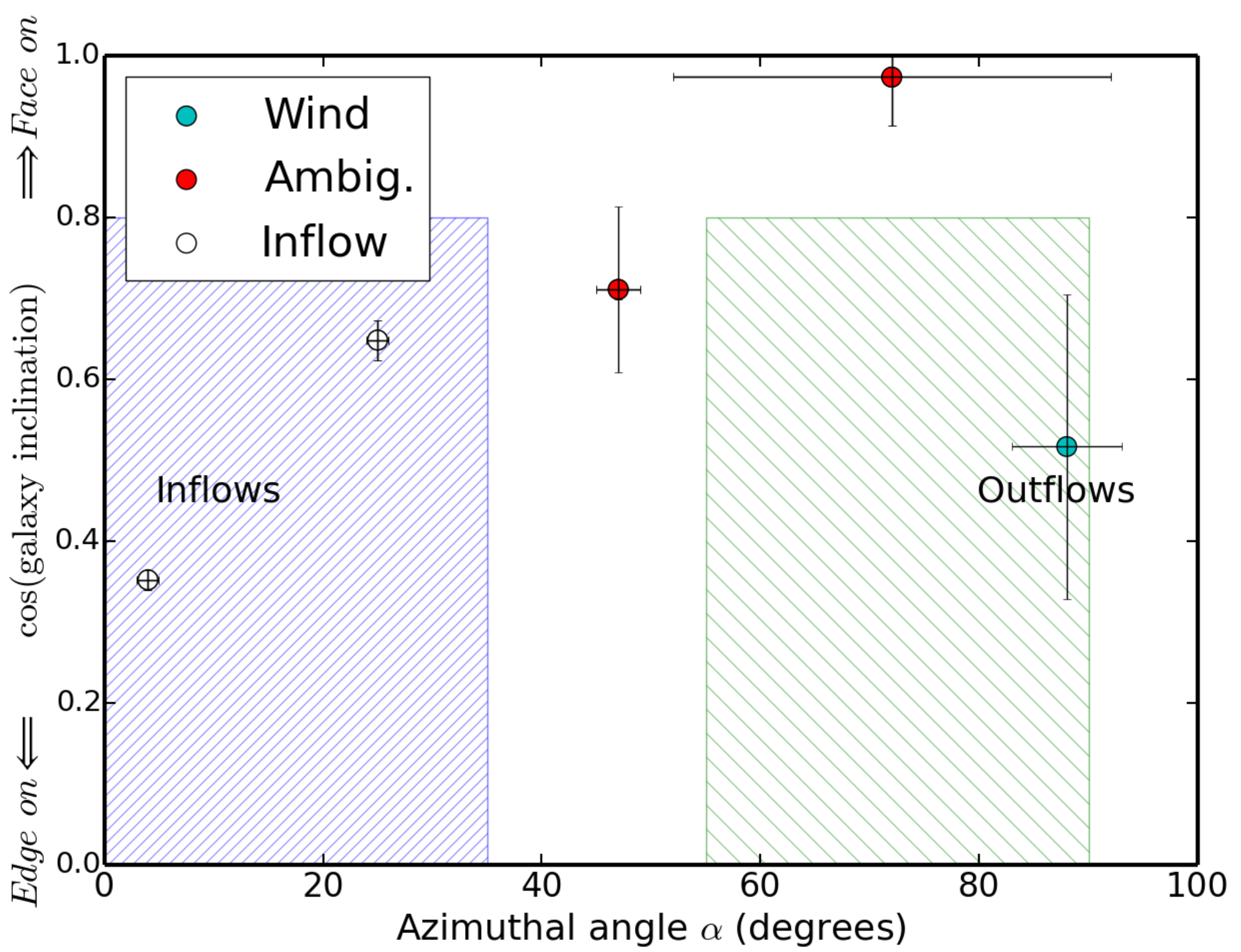}
  \caption{Galaxy inclinations as a function of azimuthal angle $\alpha$ for the 5 non-merger galaxies detected in the two fields J213748+0012 and J215200+0625. 
  We note that only one galaxy is classified as a wind-pair. 
  The dashed areas correspond to azimuthal angle ranges for which we classify pairs as inflow-pairs (blue and narrow dashes) or wind-pairs (green and wider dashes). 
  These areas stop for face-on galaxies as uncertainty on position angles are too large and thus difficult to classify pairs. 
  \label{fig:incl_vs_alpha}
}
\end{figure}

In addition to the azimuthal angle, if a galaxy has a low inclination, classification  can be ambiguous given that the uncertainty on the position angle will be large.
Figure~\ref{fig:incl_vs_alpha} shows galaxy inclination as a function of quasar azimuthal angle.
From the 5 detected galaxies in the two quasar fields that are non-mergers, 2 are classified as inflow$-$pairs, one is an ambiguous case as its azimuthal angle is 47$^\circ$,
one is a face-on galaxy and 
only 1 (\textbf{J215200+0625G2}) can be robustly classified as a wind-pair. 

\subsubsection{SDSSJ213748+0012G1 galaxy}
\begin{figure*}[]
  \centering
  \includegraphics[width=18cm]{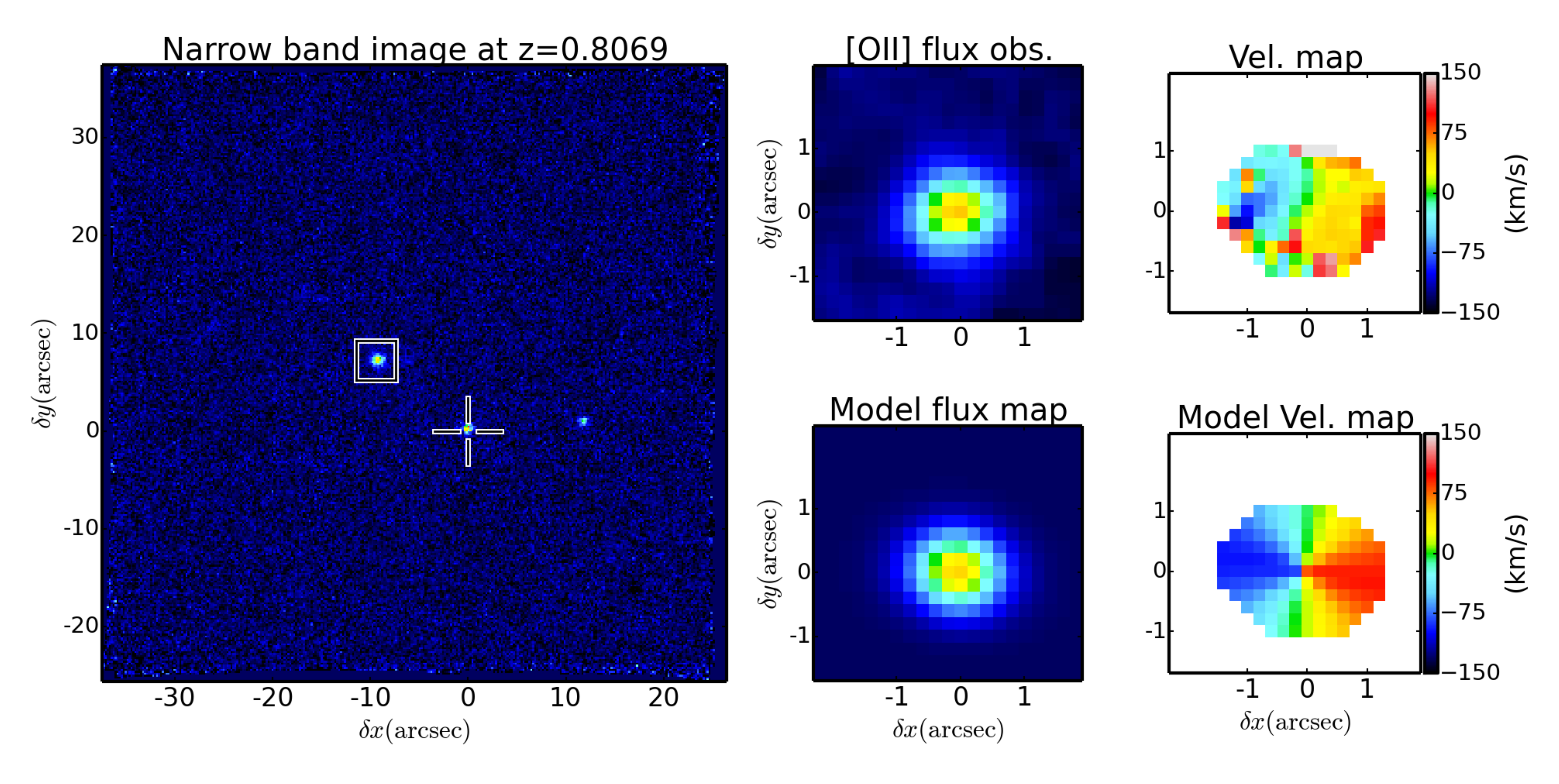}
  \caption{GalPaK and Camel results on galaxy J213748+0012G1.
  \textit{Left:} Narrow band image (30 pixels corresponding to 37.5 \AA) for [\OII] $\lambda3727, 3729$ at redshift $z=0.8069$. 
  The quasar LOS is represented by the white cross and the galaxy is inside the white rectangle. 
  The other spot on the right corresponds to continuum residuals from a star. 
  \textit{Right:} from left to right: [\OII] doublet integrated flux and velocity maps. 
The top row corresponds to a 2$\times$2 Gaussian smoothed flux map (the left panel) and Camel velocity map (top right). 
The bottom row represents the \galpak\ model flux (left) and PSF-deconvolved velocity maps (right). 
Color bars on the right show the velocities of the corresponding Velocity maps, in \kms.
This galaxy has a maximum SNR/pixel of $\approx$11.
}
  \label{fig:J2137G1}
\end{figure*}

The first detected galaxy (`G1') in the SDSSJ213748+0012 quasar field (Figure~\ref{fig:J2137G1}) has an impact parameter $b\approx 88$~kpc and corresponds to the $z_{\rm abs}\approx 0.8063$ \MgII\ 
absorption lines with a REW $W_r^{\lambda2796}$ of $0.789$~\AA. 
This J213748+0012G1 galaxy is inclined by $i\approx49\pm1.4^\circ$ and its derived maximum rotation velocity is $V_{\rm max}\approx127\pm5$~\kms. 
With an [\OII] integrated flux of $8.7\times10^{-17}$\flux, its SFR is $\approx6.3\pm0.7$~\mpy. 
In Figure~\ref{fig:J2137G1}, we can see that the morphology and the position angle is well reproduced by \galpak. 
The azimuthal angle $\alpha$ with the quasar LOS is $\alpha=25$ deg, i.e. the LOS is aligned with the major-axis.

\subsubsection{SDSSJ213748+0012G2 galaxy}
\begin{figure*}[]
  \centering
  \includegraphics[width=18cm]{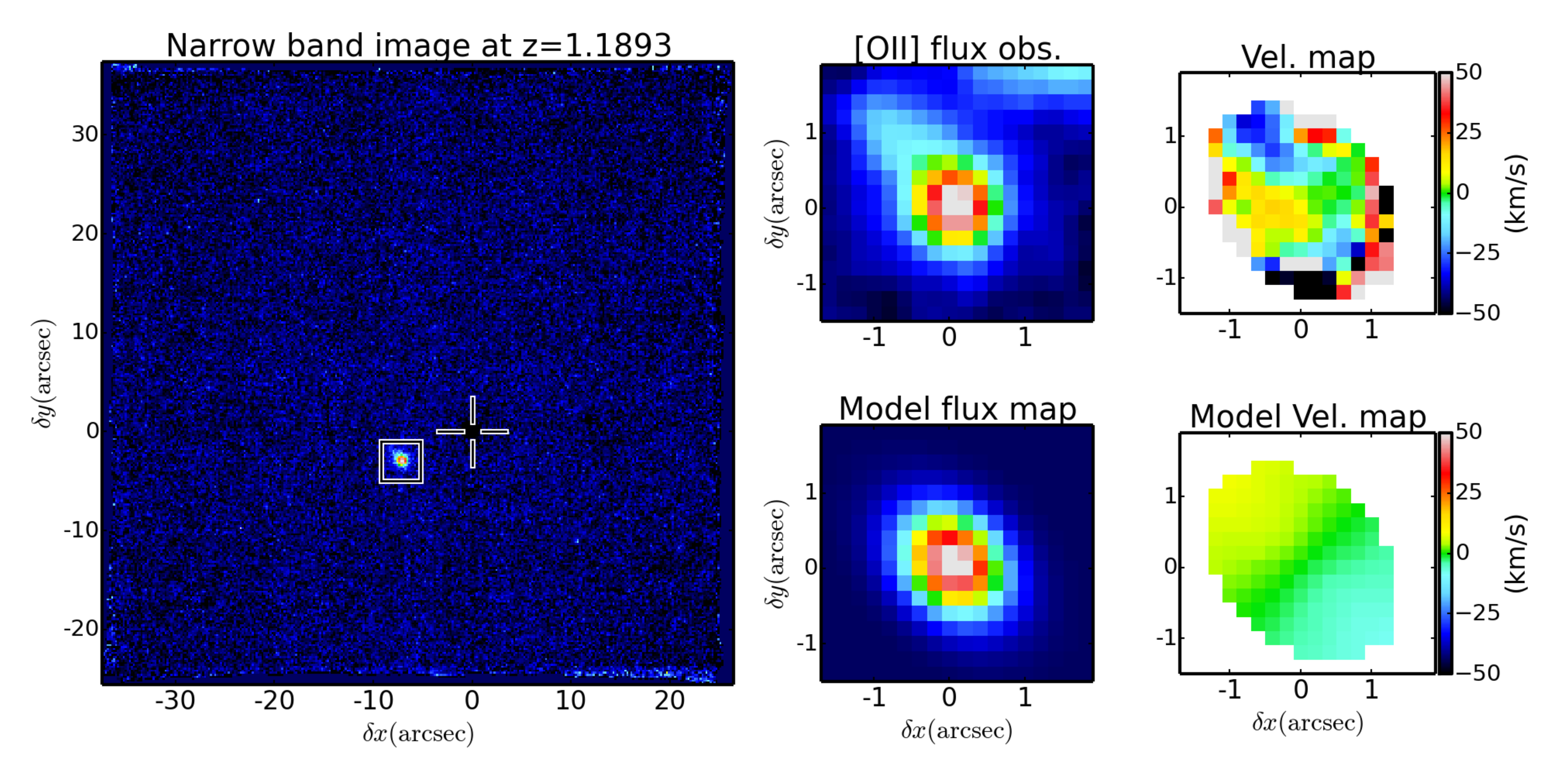}
  \caption{Same as Figure~\ref{fig:J2137G1} but for J213748+0012G2 at redshift $z=1.1893$.
  This galaxy has a maximum SNR/pixel of $\approx$11. 
  For this galaxy, we can see that the velocity maps do not agree with each other. 
  Because one part of the galaxy is not reproduced by our model and clearly has a flux component (top middle panel), 
  this galaxy seems to be a merger and therefore the azimuthal angle of this pair is not reliable.
}
  \label{fig:J2137G2}
\end{figure*}

  The galaxy J213748+0012G2 (Figure~\ref{fig:J2137G2}) 
corresponding to the $z_{\rm abs}\approx1.1890$ \MgII\ absorption lines with a REW $W_r^{\lambda2796}$ of 0.308~\AA\ 
in the J213748+0012 quasar spectrum, 
has an impact parameter of $b\approx64$~kpc and a total [\OII] doublet flux of $1.47 \times 10^{-16}$ \flux. 
From the [\OII] integrated flux we derive a SFR of $\approx41\pm8.0$~\mpy. 
This galaxy has a large velocity dispersion $\sigma\approx114\pm2.3$~\kms, i.e. it is a dispersion dominated system with $V/\sigma\sim0.2$.
Furthermore,  the velocity field derived from the line fitting algorithm Camel does not agree with its morphology, i.e.
its morphological and kinematic main axes are strongly misaligned, by $\approx80^\circ$ (Figure~\ref{fig:J2137G2}).
This  is a strong indication for a merger, and therefore this galaxy will not be considered as a wind case 
since the position angle of this galaxy is ambiguous.

\subsubsection{SDSSJ213748+0012G3 galaxy}
\begin{figure*}[]
  \centering
  \includegraphics[width=18cm]{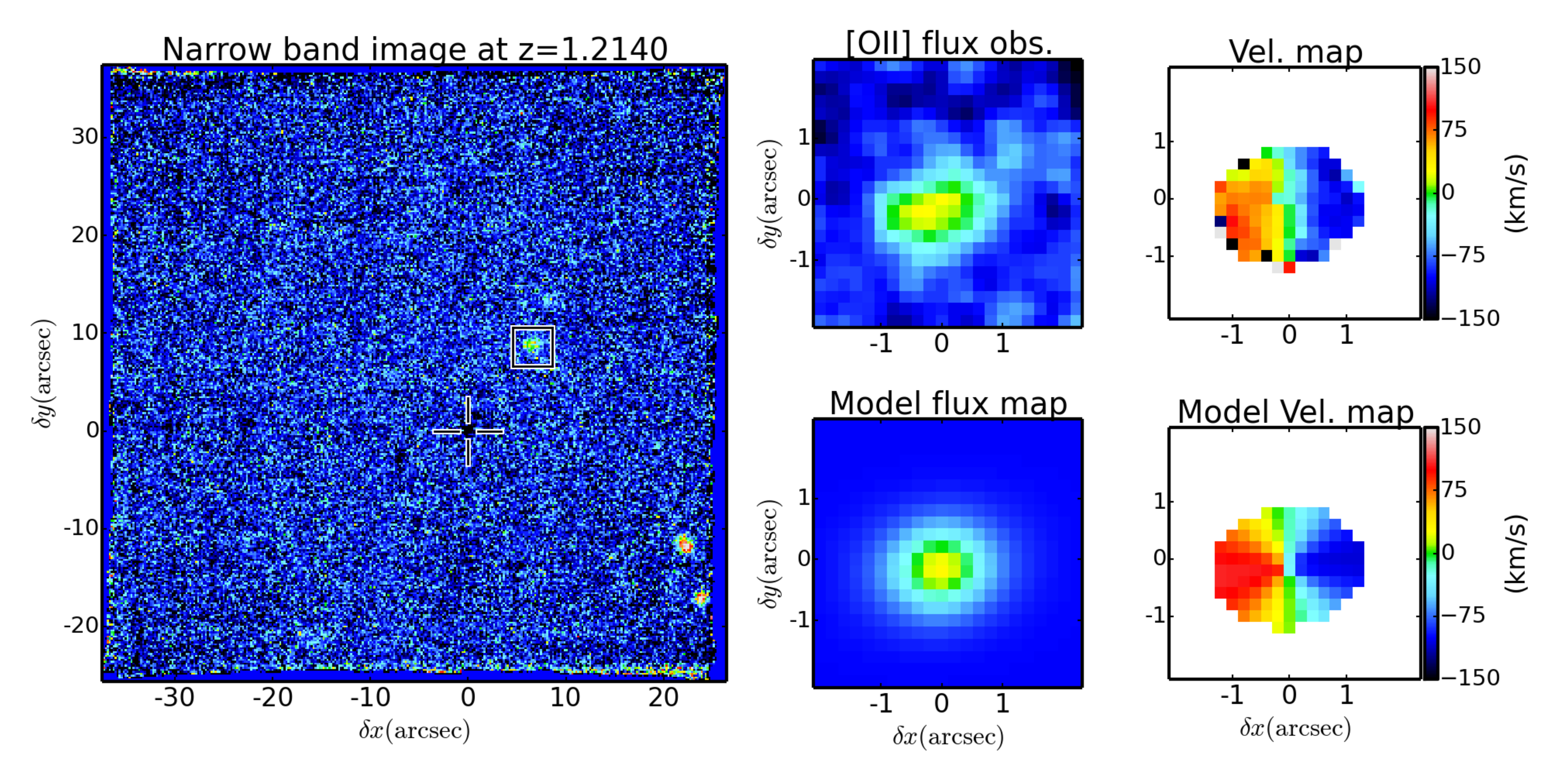}
  \caption{Same as Figure~\ref{fig:J2137G1} but for J213748+0012G3 at redshift $z=1.2140$.
  This galaxy has a maximum SNR/pixel of 4.5. 
  The spots located bottom right in the narrow band image corresponds to other galaxies. 
  These galaxy have very low probability to be the host of the \MgII\ absorption line in the quasar spectrum as they are located further away from the quasar LOS (212~kpc and 246~kpc).
}
  \label{fig:J2137G3}
\end{figure*}

 The other galaxy (J213748+0012G3, Figure~\ref{fig:J2137G3}) from the J213748+0012 field corresponding to the \MgII\ absorption lines at redshift $z_{\rm abs}\approx 1.2144$ and a REW
 $W_r^{\lambda2796}$ of 1.144\AA\ 
has an impact parameter $b$ of $\approx$ 87~kpc. 
This galaxy has an inclination $i\approx40\pm5^\circ$, a maximum rotational velocity \Vmax $\approx166\pm18$~\kms 
and an [\OII] flux of $4.17 \times 10^{-17}$~\flux. 
From this flux we derive a SFR of $\approx8.9\pm1.1$~\mpy. 
Contrary to J213748+0012G2, the kinematic and morphological PAs agree well (Figure~\ref{fig:J2137G3}), hence the 3D \galpak\ model
accounts for the 3D emission of this galaxy. 
In this case, the quasar LOS is at $\approx45^\circ$ from  the major axis of this galaxy, this pair is thus classified as ambiguous. 

\subsubsection{SDSSJ215200+0625G1 galaxy}
\begin{figure*}[]
  \centering
  \includegraphics[width=18cm]{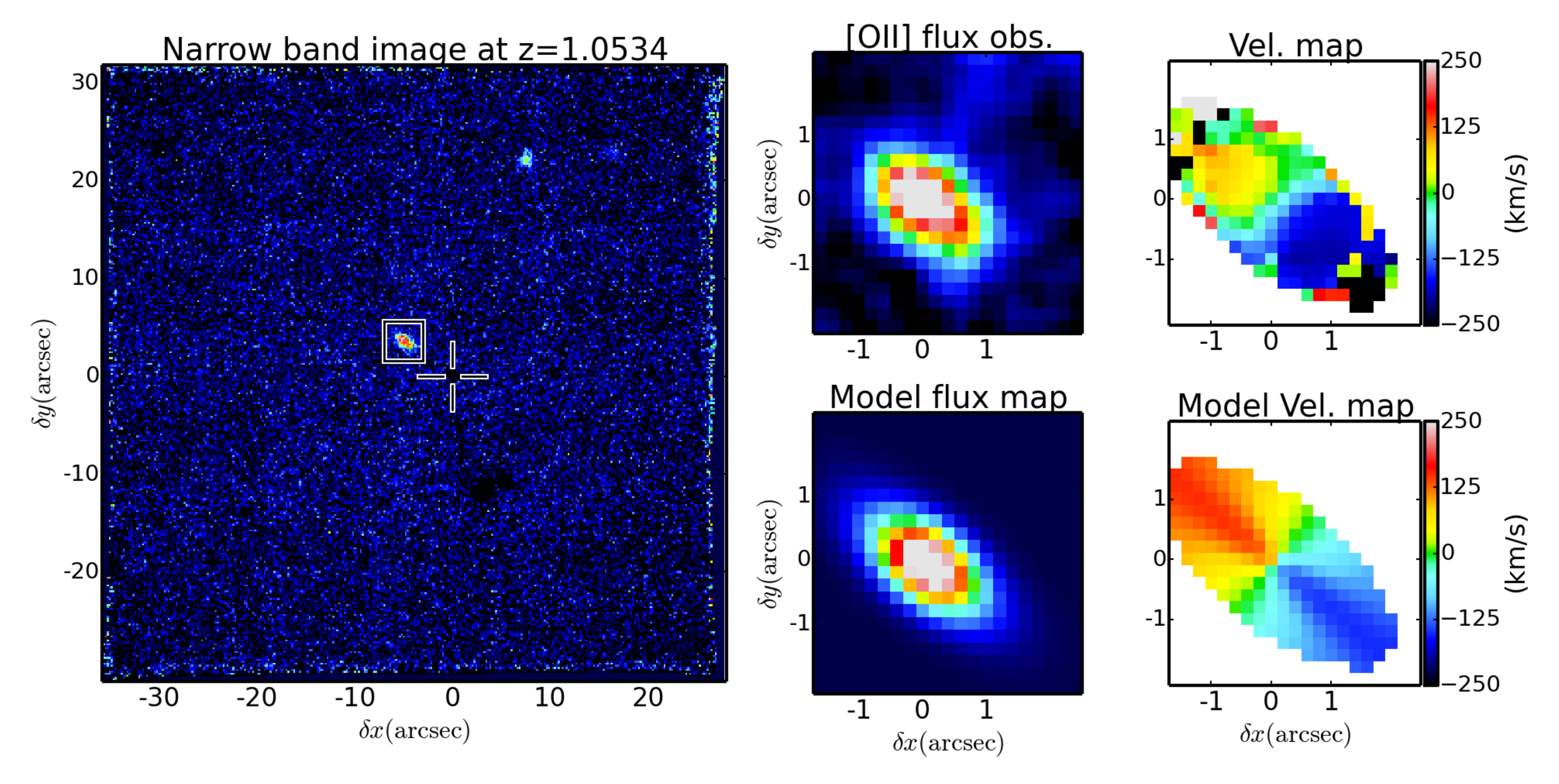}
  \caption{Same as Figure~\ref{fig:J2137G1} but for J215200+0625G1 at redshift $z=1.0534$.
  This galaxy has a maximum SNR/pixel of 9.3.
  The spot located top middle-right in the narrow band image corresponds to another galaxy. 
  Like the one in Figure~\ref{fig:J2137G3}, this galaxy is less likely to be the host of the \MgII\ absorption line in the quasar spectrum as it is located further away from the quasar LOS (189~kpc).
}
  \label{fig:J2152G1}
\end{figure*}

 The first detected galaxy from the SDSSJ215200+0625 quasar field corresponds to the \MgII\ absorption lines at redshift $z_{\rm abs}\sim1.0534$ with a REW $W_r^{\lambda2796}$ of 0.545~\AA. 
This galaxy (J215200+0625G1) has an impact parameter $b\approx45$~kpc, a maximum rotational velocity $V_{\rm max}\approx 161\pm2$~\kms\ and an inclination $i\approx69\pm0.7^\circ$. 
With an [\OII] integrated flux of $1.09\times 10^{-16}$ we derive a SFR of $\approx19.0\pm3.1$~\mpy. 
For this galaxy, Figure~\ref{fig:J2152G1} shows a good agreement between \galpak\ and Camel flux and velocity maps. 
We can clearly see that the quasar LOS is aligned with the major axis of this galaxy with $\alpha=4$ deg and is thus classified as an inflow-pair. 

\begin{figure*}[]
  \centering
  \includegraphics[width=18cm]{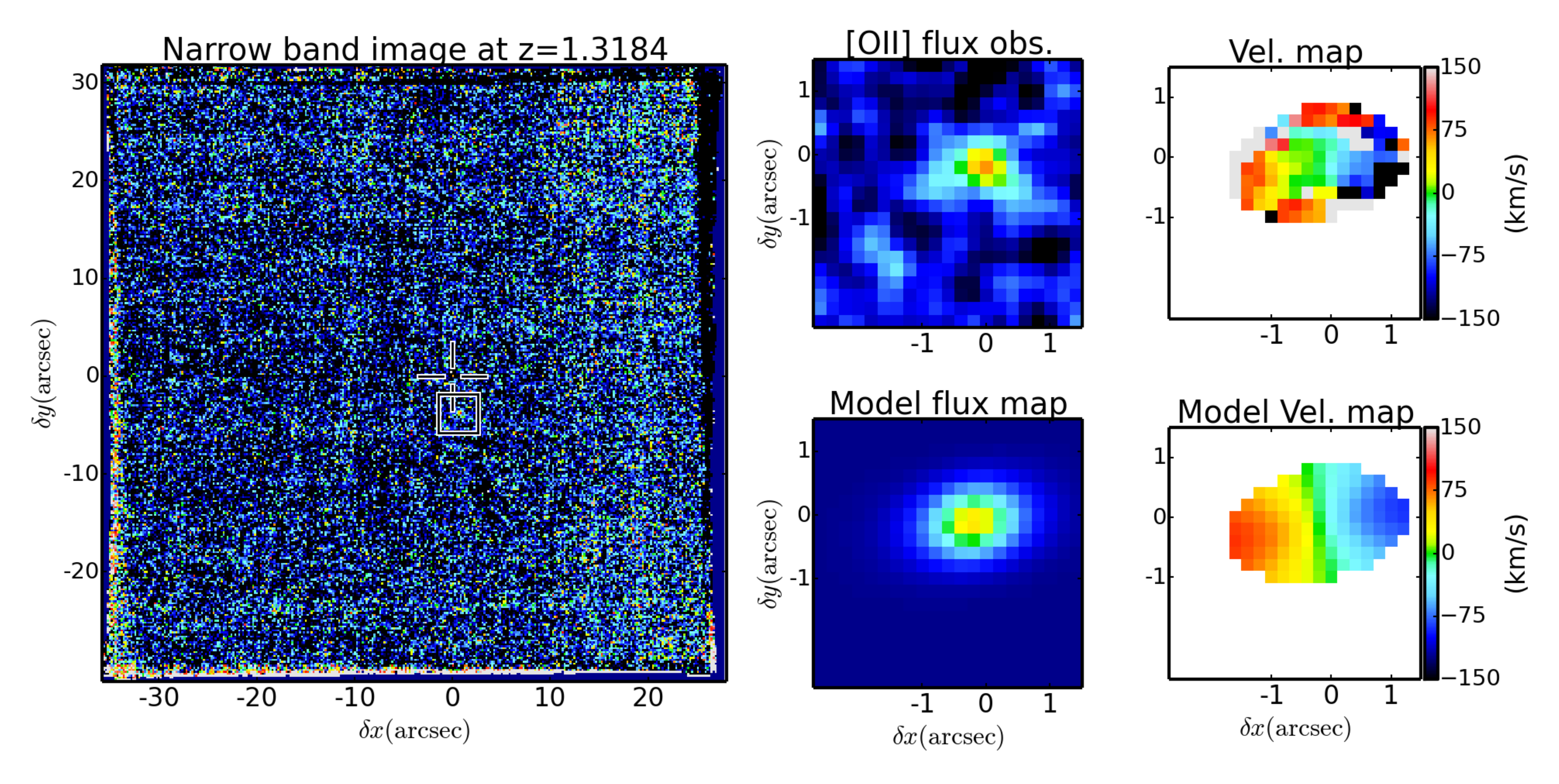}
  \caption{Same as Figure~\ref{fig:J2137G1} but for \textbf{J215200+0625G2} at redshift $z=1.3184$.
  This galaxy has a maximum SNR/pixel of 4.2 and is thus difficult to see in the left image but can be seen in the smoothed [OII] flux image.
}
  \label{fig:J2152G2}
\end{figure*}

\textbf{\subsubsection{SDSSJ215200+0625G2 galaxy}}\footnote{In all the paper (text, Tables and Figures), the only wind-pair will appear in bold font to help the reader}
 The galaxy (\textbf{J215200+0625G2}) corresponding to the redshift $z_{\rm abs}\approx1.3190$ \MgII\ absorption lines with a rest equivalent 
width $W_r^{\lambda2796}$ of 1.424~\AA\ has an impact parameter $b\approx34$~kpc.
The derived galaxy redshift is 1.31845 with an inclination of $i\approx59\pm11^\circ$ and a maximum rotational velocity \Vmax~$\approx130\pm29$~\kms.
With an [\OII] flux of $\approx 1.99 \times 10^{-17}$~\flux, we derive a SFR of $\approx4.6\pm0.4$~\mpy. 
Even if this galaxy is faint, as seen in Figure~\ref{fig:J2152G2}, its \galpak-derived morphology and position angle are in good agreement with Camel maps. 
The quasar LOS is aligned with the minor axis of this galaxy with $\alpha=88\pm5$ deg. 

\subsubsection{SDSSJ215200+0625G3 galaxy}
\label{2152G3}
\begin{figure*}[]
  \centering
  \includegraphics[width=18cm]{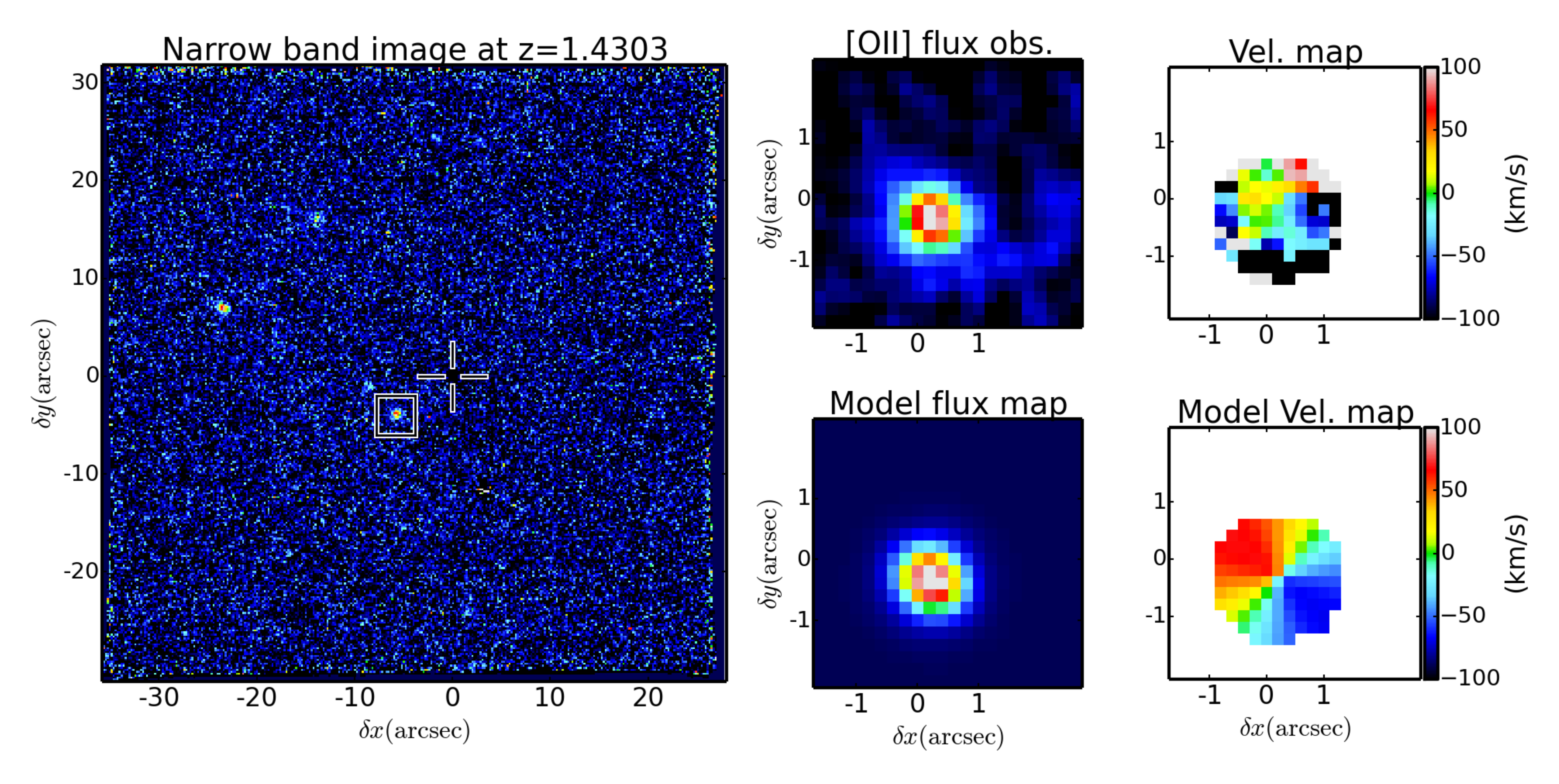}
  \caption{Same as Figure~\ref{fig:J2137G1} but for J215200+0625G3 at redshift $z=1.4303$. 
  Note that the emissions around the galaxy in the observed [\OII] flux panel is noise and not tidal tails.
  This galaxy has a maximum SNR/pixel of 10.5.
  Again, as in Figures~\ref{fig:J2137G3} and \ref{fig:J2152G1}, residual spots are galaxies further away from the quasar LOS and are thus less likely to be the host of the absorbing materials (78, 184 and 211~kpc). 
  The 78~kpc away galaxy is close enough to be considered as an host galaxy but we choose to ignore it based on impact parameter argument. 
}
  \label{fig:J2152G3}
\end{figure*}

 The last galaxy (J215200+0625G3) in the J215200+0625 quasar field has an impact parameter $b\approx63$~kpc and corresponds to the \MgII\ absorption lines at redshift $z_{\rm abs}\approx1.4309$ 
with $W_r^{\lambda2796}=1.152$~\AA. 
The galaxy has an inclination of $i\approx13\pm4^\circ$, a maximum rotational velocity \Vmax~$\approx298\pm40$~\kms\ 
and an [\OII] integrated flux of $\approx5.05 \times 10^{-17}$~\flux. 
With this flux we derive a SFR of $\approx19\pm3.0$~\mpy. 
Figure~\ref{fig:J2152G3} shows that the morphology is in agreement with Camel but the position angle derived for this galaxy is more uncertain due to the low inclination of this galaxy. 
With an azimuthal angle of $\alpha=72\pm20$ deg and its low inclination, we cannot determine whether the quasar LOS is aligned with the minor or major axis of the galaxy. 

\begin{table*}[]
\centering
\caption{Morpho-kinematics results on host galaxies. \label{table:galpak}
}
\begin{tabular}{lccclcllcll}
\hline
Galaxy 		&$z_{\rm abs}$		&$z_{\rm gal}$	&$b$	&SNR	&Size  		&$i$		&$V_{\rm max}$	&Flux			&$\alpha$	&Class	\\
(1) 		&(2)			&(3)	&(4)		&(5)	&(6)		&(7)		&(8)		&(9)			&(10)		&(11)	\\
\hline
J213748+0012G1	&0.8063		&0.80690$\pm$0.00001	&88.1$\pm$0.2	&11	&2.43$\pm$0.06	&49.6$\pm$1.4	&126.2$\pm$5	&$8.67\times10^{-17}$	&25$\pm1$	&Inflow	\\
J213748+0012G2	&1.1890		&1.18925$\pm$0.00001	&63.7$\pm$0.2	&11	&3.15$\pm$0.08	&55.6$\pm$0.8	&15.9$\pm$8	&$1.47\times10^{-16}$	& $\cdots$	& Merger	\\
J213748+0012G3	&1.2144		&1.21397$\pm$0.00003	&87.2$\pm$0.2	&4.5	&5.38$\pm$0.33	&40.4$\pm$5.0	&166.5$\pm$18	&$4.18\times10^{-17}$	&47$\pm2$	&Ambig.	\\
J215200+0625G1	&1.0534		&1.05335$\pm$0.00001	&45.4$\pm$0.2	&9.3	&5.52$\pm$0.09	&69.4$\pm$0.7	&161.4$\pm$2	&$1.09\times10^{-16}$	&4$\pm1$	&Inflow	\\
\textbf{J215200+0625G2}	&1.3190		&1.31843$\pm$0.00005	&34.0$\pm$0.2	&4.2	&3.06$\pm$0.51	&58.9$\pm$10.8	&130.6$\pm$29	&$1.99\times10^{-17}$	&88$\pm5$	&Wind	\\
J215200+0625G3	&1.4309		&1.43033$\pm$0.00004	&62.5$\pm$0.2	&10.5	&1.51$\pm$0.12	&13.3$\pm$3.4	&298.5$\pm$39	&$5.05\times10^{-17}$	&72$\pm20$	&Wind$/$Ambig.	\\
\hline
\end{tabular}\\
\vspace*{0,1cm}
{
(1) Quasar name;
(2) \MgII\ absorption redshift;
(3) Galaxy redshift;
(4) Impact parameter (kpc);
(5) SNR per pixel;
(6) Galaxy half-light radius (kpc);
(7) Galaxy inclination (degrees);
(8) Galaxy maximum velocity (\kms );
(9) Integrated [\OII] flux of the galaxy (\flux);
(10) Azimuthal angle (degrees);
(11) Class (inflow-pair/wind-pair) based on $\alpha$ selection.
}
\end{table*}


\subsection{Radial dependence of CGM}
\label{REW}

\begin{figure*}[h!]
  \centering
  \includegraphics[width=18cm]{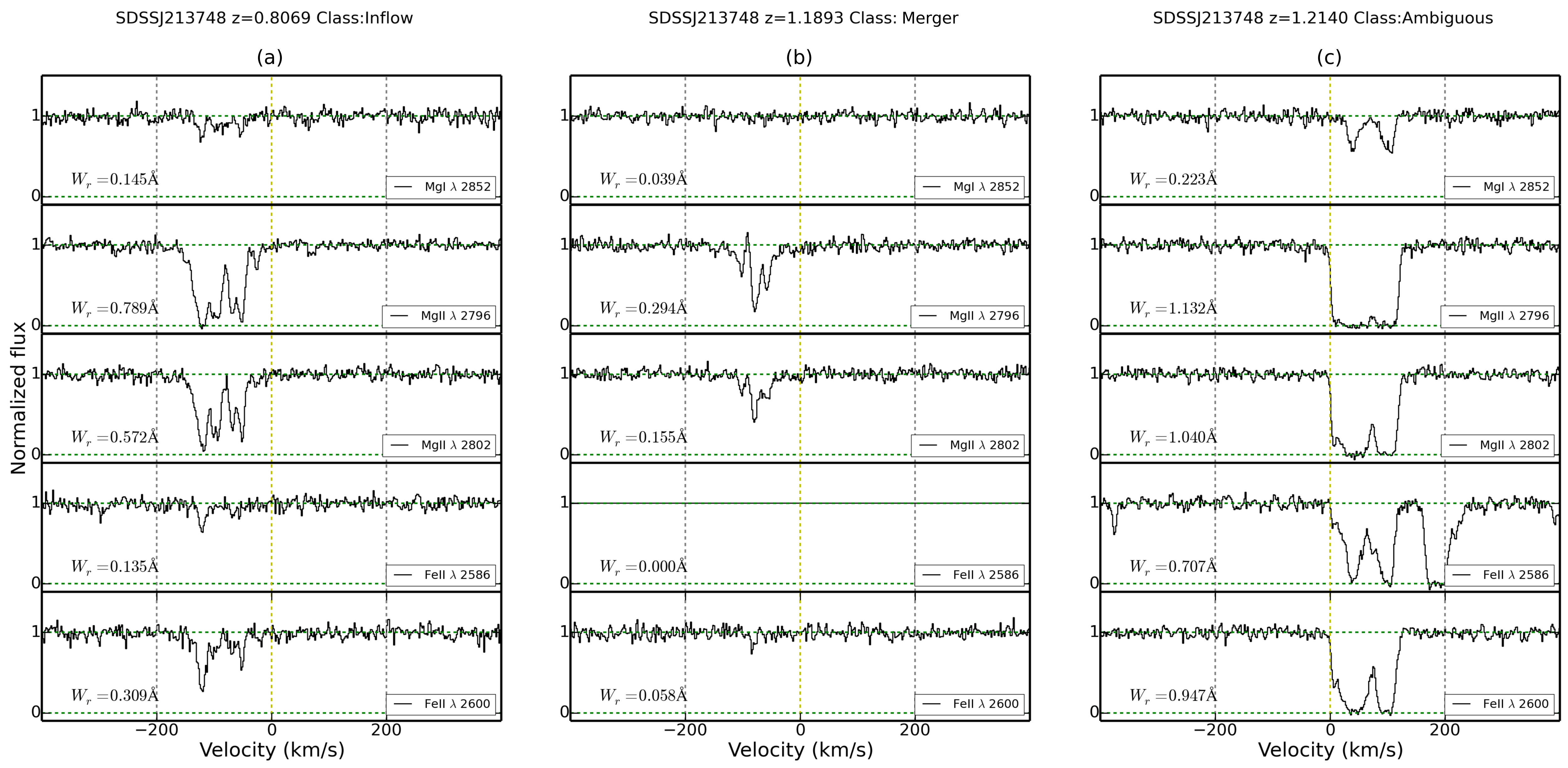}
  \caption{UVES \MgI\ \mc, \MgII\ $\lambda \lambda 2796,2802$ and \FeII\ $\lambda \lambda 2586,2600$ absorption lines centered 
  at host galaxy systemic velocity for the SDSSJ213748+0012 quasar spectrum. 
  The left panel corresponds to absorption lines from the SDSSJ213748+0012G1 host galaxy (a). 
  The middle panel to the SDSSJ213748+0012G2 host galaxy (b), 
  and right panel to SDSSJ213748+0012G3 (c). 
  Note that in the right column, the \FeII\ $\lambda 2586$ REW is calculated without the $\approx$200~\kms\ absorption component.}
  \label{fig:J213748_abs}
\end{figure*}

\begin{figure*}[h!]
  \centering
  \includegraphics[width=18cm]{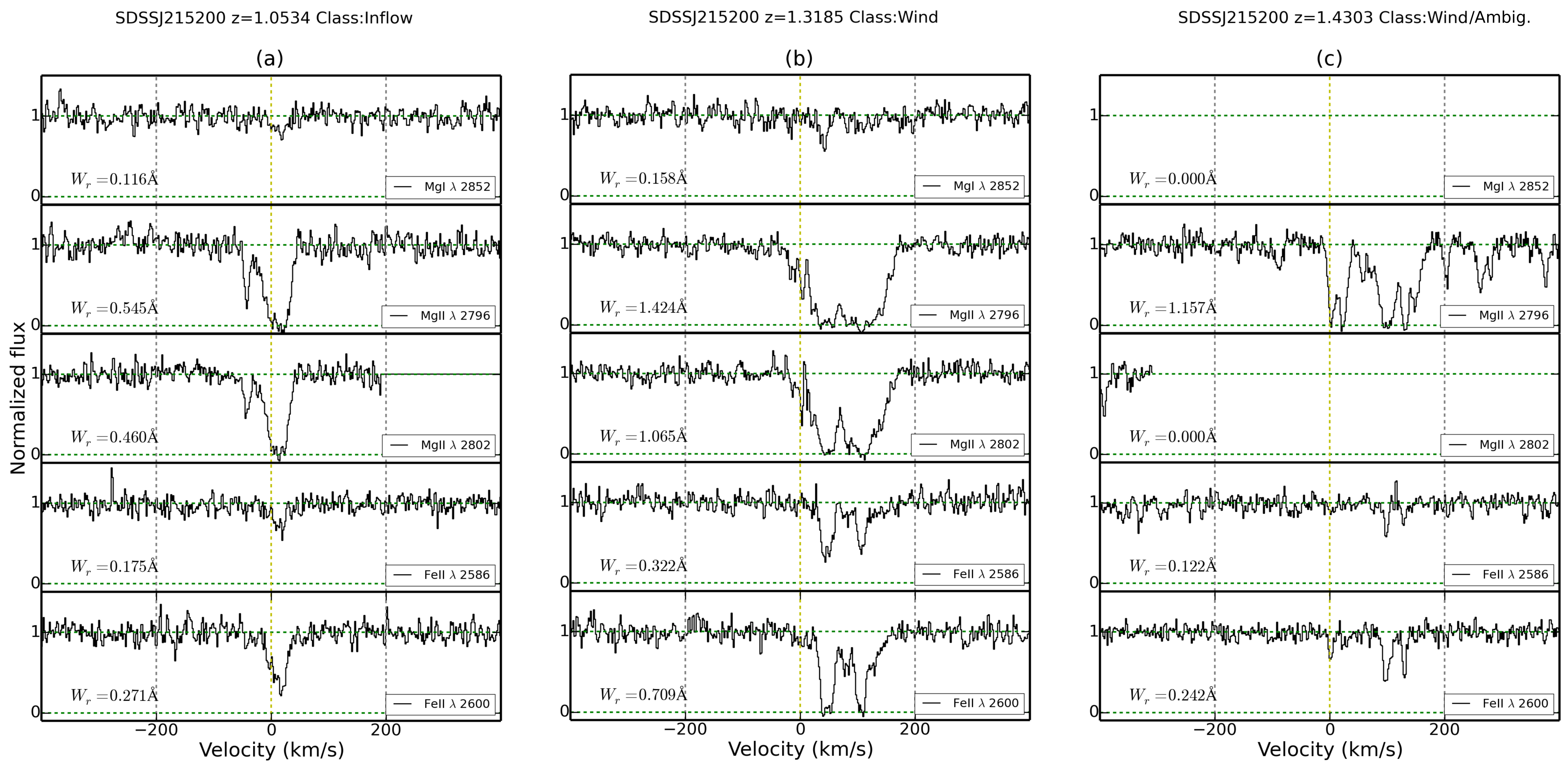}
  \caption{UVES \MgI\ \mc, \MgII\ $\lambda \lambda 2796,2802$ and \FeII\ $\lambda \lambda 2586,2600$ absorption lines centered 
  at host galaxy systemic velocity for the SDSSJ215200+0625 quasar spectrum. 
  The left panel corresponds to absorption lines from the SDSSJ215200+0625G1 host galaxy (a). 
  The middle panel to the \textbf{SDSSJ215200+0625G2} host galaxy (b), 
  and right panel to SDSSJ215200+0625G3 (c).}
  \label{fig:J215200_abs}
\end{figure*}

For each quasar spectrum, we measure the REW for the \MgII\ absorption lines ($W_{r}^{\lambda 2796}$) in the UVES data and compare them with 
the SDSS values $W_{r}^{\lambda 2796}$ (see Table~\ref{table:REW}).
We find that the results are consistent with each other.
We also calculate REWs of the \MgII\ \mb, \MgI\ \mc, \FeII\ $\lambda2586$ and \FeII\ $\lambda2600$ in UVES quasar spectra. 
Results are shown in Table~\ref{table:REW}.
Figures~\ref{fig:J213748_abs} and~\ref{fig:J215200_abs} show the UVES MgI\ \mc, \MgII\ $\lambda \lambda 2796,2802$ and \FeII\ $\lambda \lambda 2586,2600$ absorption profiles 
and label the measured REW of each profile for both quasar fields. 

One of the first deductions we can make from Figures~\ref{fig:J213748_abs} and~\ref{fig:J215200_abs} is that there is no clear difference (like different asymmetry behavior for instance) between what seems 
to be outflowing material and circum-galactic or inflowing gas concerning the different absorption lines.

\begin{figure}[]
  \centering
  \includegraphics[width=8.0cm]{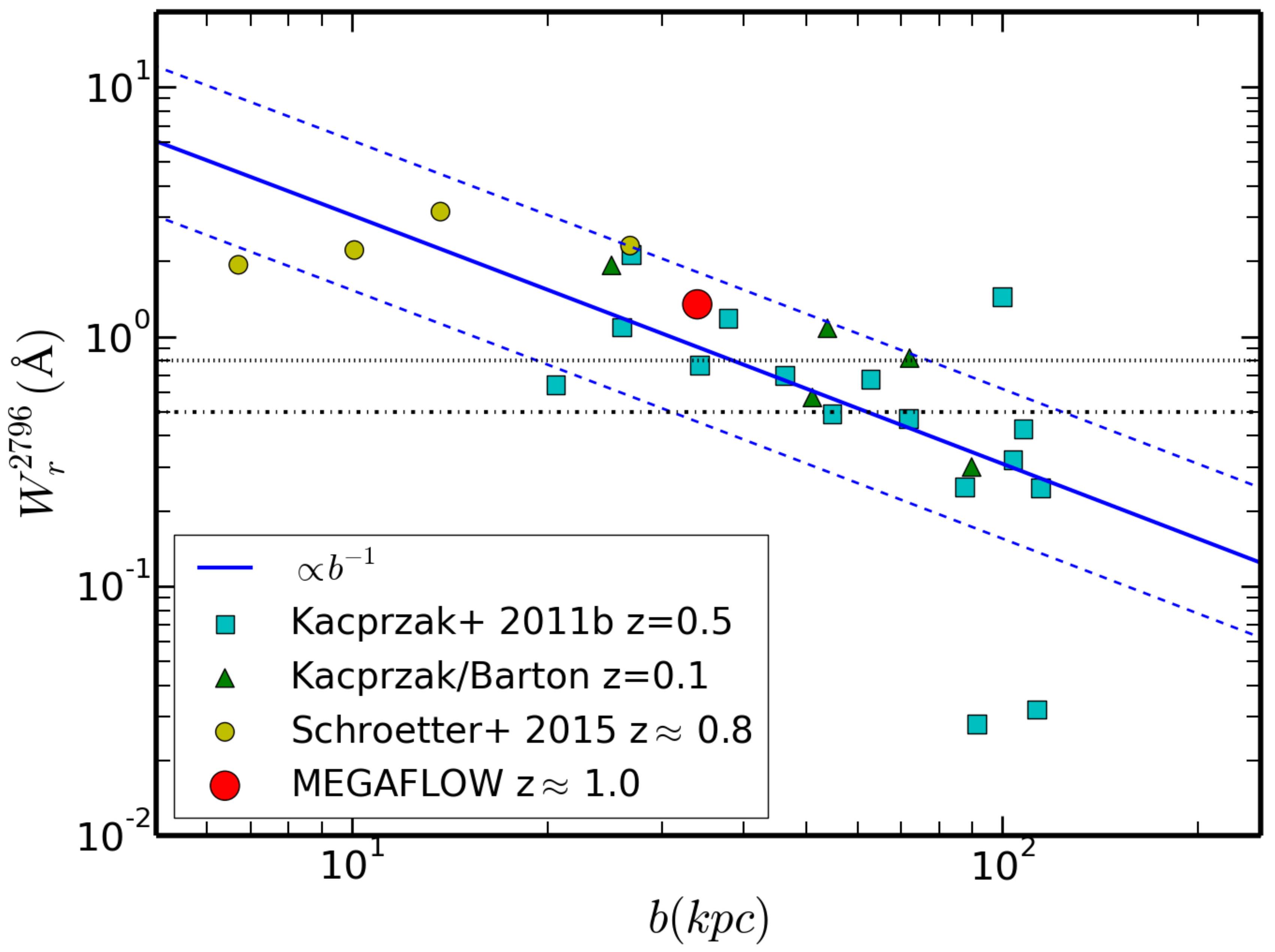}
  \caption{$W_r^{\lambda 2796}$ as a function of impact parameter $b$ for galaxy-quasar pairs classified as wind-pairs.
  The dashed blue lines show the 0.3 dex scatter. 
  The horizontal dotted black lines represent the $W_r^{\lambda 2796}$=0.8~\AA\ and $W_r^{\lambda 2796}$=0.5~\AA\ selection limits. 
  }
  \label{fig:rew_vs_b}
\end{figure}

Figure~\ref{fig:rew_vs_b} shows the distribution of REW $W^{\lambda 2796}_{\rm r}$ for pairs 
with an azimuthal angle $\alpha > 45^\circ$ as a function of impact parameter $b$ for this work 
as well as \citet{kacprzak_11b, kacprzak_11} and \citet{schroetter_15}.
This Figure shows that for wind pairs, as mentioned in \citet{bouche_12}, we clearly see a tight correlation between $W^{\lambda 2796}_{\rm r}$ and $b$.
This $W^{\lambda 2796}_{\rm r} - b$ correlation goes approximatively as $b^{-1}$. 
This figure shows that the anti-correlation between impact parameter $b$ and $W_r$ is again confirmed at $b<100$ kpc. 
The scatter around the relation in Figure~\ref{fig:rew_vs_b} is $\approx0.3$ dex (delineated with the dotted lines). 
The solid line traces the fiducial $1/b$ relation for mass-conserved bi-conical outflows \citep[see][]{bouche_12}. 

\begin{table*}[h!]
\centering
\caption{UVES rest equivalent widths. \label{table:REW}
}
\begin{tabular}{lcccccccl}
\hline
Galaxy 		&$W_r^{\lambda \rm 2796}$(SDSS)	&$W_r^{\lambda \rm 2796}$	&$W_r^{\lambda \rm 2802}$	&$W_r^{\lambda \rm 2852}$	&$W_r^{\lambda \rm 2586}$	&$W_r^{\lambda \rm 2600}$	&$log(N_{\HI})$	&Class	\\
(1) 		&(2)				&(3)				&(4)				&(5)				&(6)				&(7)				&(8)		&(9)\\
\hline
J213748+0012G1	&0.724$\pm$0.09			&0.789$\pm$0.02			&0.572$\pm$0.02			&0.145$\pm$0.02			&0.135$\pm$0.02			&0.309$\pm$0.02			&19.24		&Inflow	\\
J213748+0012G2	&0.308$\pm$0.06			&0.294$\pm$0.02			&0.155$\pm$0.02			&0.039$\pm$0.02			&$\cdots$			&0.058$\pm$0.02			&18.61		& Merger	\\
J213748+0012G3	&1.122$\pm$0.06			&1.132$\pm$0.02			&1.040$\pm$0.02			&0.223$\pm$0.02			&0.707$\pm$0.02			&0.947$\pm$0.02			&19.58		&Ambig.	\\
J215200+0625G1	&0.522$\pm$0.14			&0.545$\pm$0.02			&0.460$\pm$0.02			&0.116$\pm$0.02			&0.175$\pm$0.02			&0.271$\pm$0.02			&19.01		&Inflow	\\
\textbf{J215200+0625G2}	&1.347$\pm$0.12			&1.424$\pm$0.02			&1.065$\pm$0.02			&0.158$\pm$0.02			&0.322$\pm$0.02			&0.709$\pm$0.02			&19.71		&Wind	\\
J215200+0625G3	&1.152$\pm$0.11			&1.157$\pm$0.02			&$\cdots$			&$\cdots$			&0.122$\pm$0.02			&0.242$\pm$0.02			&19.59		& Wind$/$Ambig.	\\
\hline
\end{tabular}\\
\vspace*{0,1cm}
{
(1) Quasar name;
(2) SDSS \MgII\ \ma\ rest equivalent width (\AA );
(3) UVES \MgII\ \ma\ rest equivalent width (\AA );
(4) UVES \MgII\ \mb\ rest equivalent width (\AA );
(5) UVES \MgI\ \mc\ rest equivalent width (\AA );
(6) UVES \FeII\ $\lambda2586$ rest equivalent width (\AA );
(7) UVES \FeII\ $\lambda2600$ rest equivalent width (\AA );
(8) Gas column density at the impact parameter (cm$^{-2}$);
(9) Class (inflow-pair/wind-pair) based on $\alpha$ selection.
}
\end{table*}

\section{Wind model}
\label{wind}

In this section, we describe the wind modeling. 
We create a cone having an opening angle corresponding to \thetam\footnote{\thetam is defined from the central axis, 
and the cone subtends an area $\Sigma$ of $\pi \cdot \theta_{\rm max}^2$.} 
and fill it randomly with particles representing cold gas clouds being pushed away by a hot medium or radiation pressure. 
These particles are distributed such that their number goes like $1 / r^2$, where $r$ is the distance to the galaxy center. 
The particle density is normalized arbitrarily to reproduce the optical depth of the absorption profiles. 

Such entrained clouds are accelerated to their terminal velocity quickly in a few kpc or $<10$ kpc since the pressure from
 the hot medium or the radiation field scales as $1/r^2$. The range of impact parameters for the galaxy$-$quasar pair 
in our sample is always larger than 30~kpc. Hence, we assume, for simplicity, that the particles have a constant radial velocity corresponding to \Vout. 
In addition, a single LOS probes a rather small range of distances from the host galaxy such that a gradient in the outflow velocity would have no significant impact on our results.
So far, only in one LOS with an impact parameter less than 10 kpc in \citet{schroetter_15}, we required an accelerated wind profile.

We then orient the cone following the galaxy inclination and simulate the quasar LOS such that 
the galaxy$-$quasar pair matches the geometrical configuration of the MUSE data. 

The particle velocities are then projected along the simulated quasar LOS and the distribution of the projected velocities gives us a
simulated optical depth $\tau_{\rm v}$, which we turn into an absorption profile $\propto \exp(-\tau_{\rm v})$.
In order to facilitate comparison with the data, Poisson noise is added to the simulated absorption profile to simulate the instrumental noise. 
This noise is chosen to have the same level as the data. 

The model has two main free parameters, the wind speed \Vout\ and \thetam\ the wind opening angle. 
These two parameters are independent for a given galaxy inclination as one can see from the following arguments (see also \citet{schroetter_15} for more details). 
The outer edges of the absorption profile (reddest for a cone pointing away from the observer, bluest for a cone pointing towards the observer) depends directly on the wind velocity (Figure A-1 in \citet{schroetter_15}). 
The inner edge (towards Vsys) of the absorption profile depends directly on the wind opening angle \thetam\ (Figure A-1 in \citet{schroetter_15}). 
Note that the galaxy inclination impacts the absorption profiles similarly to the \thetam\ parameter but since the inclination is determined by our 3D fit with \galpak, there are no degeneracies. 

In order to determine which model best reproduces the data, the best fit model is found by eye.
However, given that there are stochastic features in the simulated profiles,  we generate dozens of simulated profiles for a given set of parameters.
The errors on these parameters are given by the range of values allowed by the data. 
We proceed as follow: 
We first generate models changing only one parameter to fit one part of the absorption profile (outer part for \Vout\ or inner part for \thetam). 
Then, we change only the other parameter (\thetam\ or \Vout) generating other models to fit the other part of the absorption. 
We generate models with range of values of 10 to 500~\kms\ (with steps of 10~\kms) for \Vout\ and 20 to 50$^\circ$ (with steps of 5$^\circ$) for \thetam. 
As mentioned before, these two parameters being independent, there is no degeneracy between generated models. 
We use these parameters range to fit the data since outflows are likely to be collimated in a cone with an opening angle around 30$^\circ$ 
\citep[e.g.][]{chen_10, bouche_12, bordoloi_11, kacprzak_12, martin_12, bordoloi_14, rubin_14}

Examples on how the wind model behaves as we change the different parameters can be seen in the appendix of \citet{schroetter_15}. 


\subsection{The wind$-$pair case of \textbf{J215200+0625G2}}
\label{individual}

Figure~\ref{fig:J215200_abs}, middle column (b), shows the UVES \MgI\ \mc, \MgII\ $\lambda \lambda 2796,2802$ 
and \FeII\ $\lambda \lambda 2586,2600$ absorption lines for this galaxy-quasar pair. 
From this Figure, we can see that the \MgII\ $\lambda \lambda 2796,2802$ absorption lines are saturated and thus the need to simulate the absorption from \FeII\ $\lambda 2586$ which is 
the only non-saturated absorption lines in the presented transitions.

\begin{figure*}[h!]
  \centering
  \includegraphics[width=16cm]{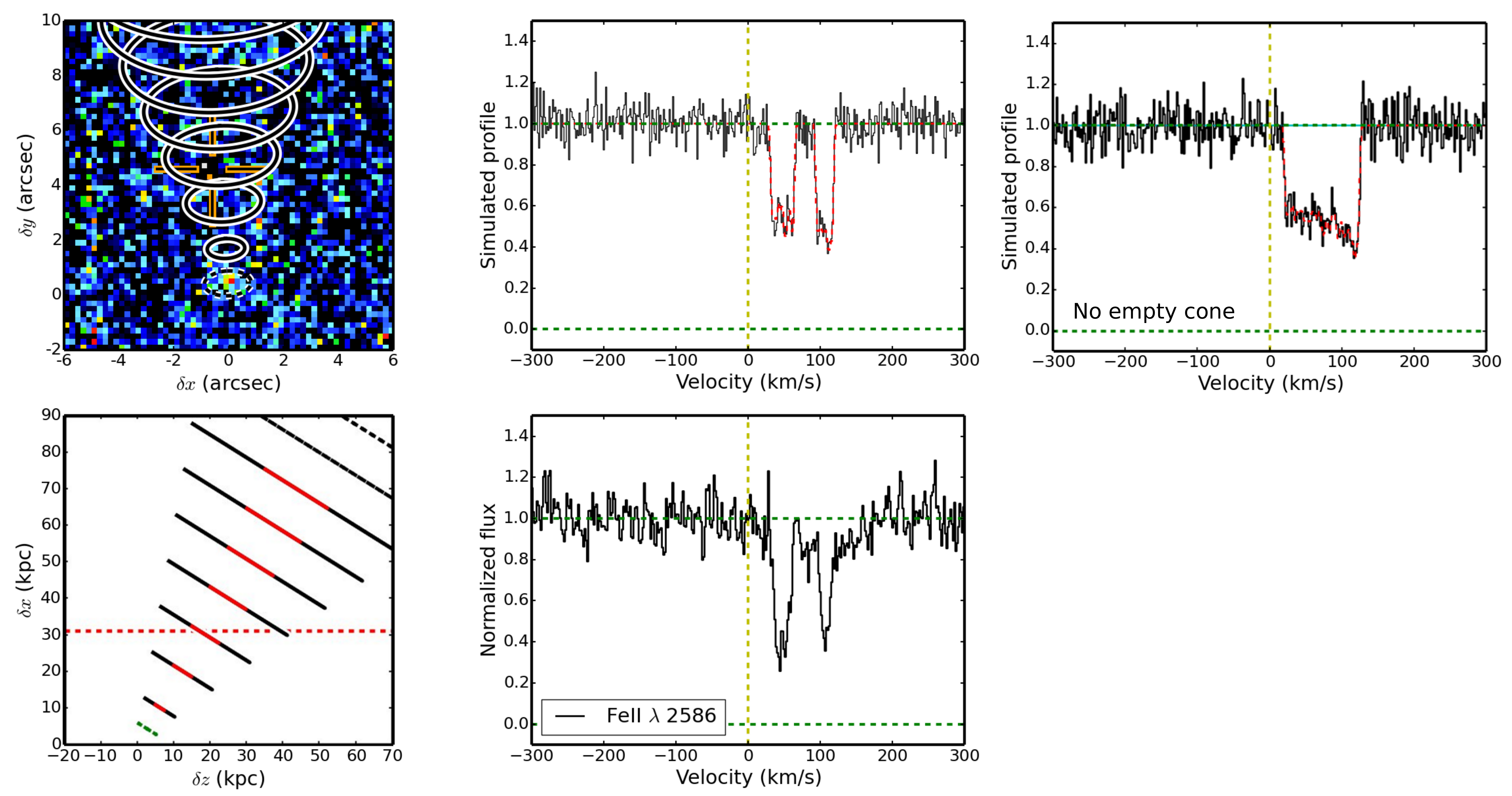}
  \caption{
  Representation of the cone model and quasar spectrum associated with the \textbf{J215200+0625G2} galaxy ($z=1.31845$). 
  \textit{Top left: }the cone model seen in the sky plane (xy). 
  This is a narrow band image centered around the galaxy [\OII] emission lines with the continuum subtracted.
  The dashed circle represents the inclined galaxy disk and the black and white inclined circles illustrate the gas outflow cone.
  The orange cross represents the position of the quasar LOS.
  \textit{Bottom left: }a side view of the cone where the z-axis corresponds to the quasar LOS direction with the observer to the left. 
  The red part of the cone represents the empty inner cone. 
  \textit{Middle: }Normalized flux for the \FeII\ ($\lambda 2586$) absorption line observed with UVES (bottom) and the reconstructed profile (top).
  Note that this model does not reproduce the depth of the absorption line. 
  In UVES simulated absorption profile, the red line corresponds to the simulated profile without any instrumental noise. 
  This wind model uses a very low-density inner cone as described in \S~\ref{individual}.
  \textit{Right: }same as the top middle panel but with no empty inner cone model. 
  This simulated profile has the same parameters as the empty inner cone one.
  We can clearly see that we cannot reproduce the gap in absorptions seen in the UVES absorption profile without the empty cone model. 
  This outflow has a \Vout\ of 150$\pm$10~\kms, a cone opening angle \thetam\ of 20$\pm$5$^\circ$ and an inner opening angle $\theta_{in}$ of 7$\pm$2$^\circ$.}
  \label{fig:J215200G2}
\end{figure*}

The bottom right panel of Figure~\ref{fig:J215200G2} shows the UVES \FeII\ $\lambda2586$ absorption lines corresponding to the \textbf{J215200+0625G2} galaxy redshift of $z=1.3184$. 
This absorption is the one we intend to fit in order to constrain outflow properties since other absorption lines like \MgII\ are saturated (see panel (b) of Figure~\ref{fig:J215200_abs}). 
In this profile, we can see a suppression of absorption around 80~\kms. 
We first tried to fit this absorption with our wind model described in \S~\ref{wind} but failed to reproduce this gap, even with stochastic effects. 
This lack of absorbing particles at these velocities shows that the outflowing cone must have a low density region inside it. 

Given that the geometry of this galaxy-quasar system (with a galaxy inclination $i$ of 59$^\circ$) and that the quasar line of sight is crossing the outflowing cone near its middle ($\alpha=88^\circ$), 
we thus developed a partially empty cone model in order to reproduce the absorption profile. 

The principle is the same as the wind model described in \S~\ref{wind} except that we only fill the cone with particles from a certain opening angle $\theta_{\rm in}$ to \thetam. 
The inner cone is thus empty. 
This model should only work if the azimuthal angle $\alpha$ of a galaxy$-$quasar system is above $\sim80^\circ$, so the quasar LOS is crossing this empty region 
and thus creating a gap of velocities in the simulated profile. 

This empty inner cone could be the signature of a hotter gas filling the inner cone while the ionized gas traced by our low-ionization lines
would correspond to the walls of the outflowing cone in a manner similar to \citet{fox_15} for the MilkyWay and to \citet{veilleux_02} for NGC1482.

Figure~\ref{fig:J215200G2} illustrates the resulting wind modeling for this galaxy. 
The first left column corresponds to the wind model representation. 
The top left panel shows a [\OII] integrated flux, continuum subtracted, image with the orange cross showing the quasar LOS position. 
The inclined circles represent the outflowing cone. 
The bottom left panel represents a side view of the cone, the quasar LOS being represented by the dashed red line, the observer being on the left. 
This representation allows us to see if the outflowing material is ejected toward or away from us, assuming our cone model is representative.
The red part of the cone represents the empty inner part.

On the middle column are represented the simulated profiles (top) and UVES spectrum around the absorption line \FeII\ $\lambda2586$ (bottom).
The red part of the simulated profile is the profile without instrumental noise and the apparent noise is due to stochastic 
effects from the Monte Carlo particle distribution. 
The red simulated absorption profile does not change much for the UVES data as compare to the noise-added one. 
We also present in Figure~\ref{fig:J215200G2}, top right panel, a similar simulated profile (with the same parameters) but without the empty inner cone model. 
We clearly see on this Figure that we cannot reproduce the gap shown in the data without an empty region. 

The bottom middle panel corresponds to UVES data. 
It corresponds to the QSO spectrum absorption lines centered at the galaxy systemic velocity. 
The element \FeII\ $\lambda2586$ corresponding to the absorption lines is shown in the bottom middle column panel.

To reproduce the shape of this absorption profile and generate the simulated profile shown in the top middle panel of Figure~\ref{fig:J215200G2}, we adjust the outflow speed \Vout\ and the cone opening angle \thetam\ 
while keeping the geometrical parameters of the galaxy fixed as described in \S~\ref{wind}. 

The best values for reproducing the UVES \FeII\ $\lambda2586$ absorption profile are an outflow velocity \Vout\ of $150 \pm 10$~\kms\ and 
a cone opening angle \thetam\ of $20\pm5^\circ$. 
For this specific case, we derive an inner opening angle of the cone of $\theta_{\rm in} \approx 7^\circ$. 

\label{J215200G2}


\subsection{Outflow rates}
\label{outflow}
Having constrained the outflow velocity and cone opening angle for the wind-pair, we can now derive the ejected mass rate $\dot M_{\rm out}$ as well as the loading factor. 

For our wind-pair, the equivalent width of the absorption lines only depends on \thetam\ and \Vout\ (see \S~\ref{wind}). 
After testing several opening angles and outflow velocities, we fitted the width of the absorption profile created by gas outflowing from the galaxy. 
The asymmetry of the profile depends on the system geometry. 
To constrain the ejected mass rate probed by the quasar LOS, we use relation~\ref{eq:Mout} from \citet{bouche_12} and \citet{schroetter_15}: 
\begin{eqnarray}
 \dot M_{\rm out} &\approx& \mu \cdot N_{\rm H}(b) \cdot b \cdot V_{\rm out} \cdot \frac{\pi}{2} \cdot \theta_{\rm max}\label{eq:Mout}\\
{\dot M_{\rm out} \over 0.5\/ \mpy\ }& \approx &{\mu \over 1.5} \cdot {N_{\rm H}(b)\over 10^{19} \rm cm^{-2}} \cdot {b\over 25 \rm kpc} \cdot {V_{\rm out} \over 200 \kms\ } \cdot {\theta_{\rm max} \over 30^{\circ}}\nn
\end{eqnarray}
$\mu$ being the mean atomic weight, $b$ the impact parameter, \thetam\ the cone opening angle\footnote{We remind the reader that \thetam is defined from the central axis, 
and the cone subtends an area $\Sigma$ of $\pi \cdot \theta_{\rm max}^2$.}, \Vout\ the outflow velocity and $N_{\rm H}(b)$ is the gas column density 
at the $b$ distance. 

The only parameter which is yet to be constrained is the gas column density $N_{\rm H}(b)$. 
To do that, we use the empirical relation~\ref{eq:menard} between the neutral gas column density 
and the \MgII\ \ma\ REW $W_r^{\lambda 2796}$ \citep{menard_09}: 

\begin{equation}
 \log (N_{\rm HI}) (\hbox{cm}^{-2}) = \log [(3.06 \pm 0.55) \times 10^{19} \times (W_r^{\lambda 2796} )^{1.7 \pm 0.26}]\label{eq:menard}.
\end{equation}

To compute the errors, we assume a gaussian error distribution. 
As described in \citet{schroetter_15}, for regions with \HI\ column density above $\log(N_{\rm HI})=19.5$, the ionized gas contribution is negligible. 
Also argued by \citet{jenkins_09}, if this column density is above this limit, one can use the correlation between \MgII\ equivalent width and $N_{\rm HI}$ as a proxy for the $N_{\rm H}$ gas column density. 
For the wind-pair \textbf{J215200+0625G2}, we have a gas column density of $\log (N_{\rm HI})\approx19.7\pm0.07$. 

Another aspect of outflow properties is whether the outflowing gas is able to escape from the galaxy gravitational well. 
To determine this, we derive the escape velocity $V_{\rm esc}$ for the \textbf{J215200+0625G2} galaxy. 
The escape velocity for an isothermal sphere is defined by Eq.~\ref{eq:Vesc} \citep{VeilleuxS_05a}.
\begin{equation}
 V_{\rm esc}=V_{\rm max}\cdot\sqrt{2\left[1+\ln \left(\frac{R_{\rm vir}}{r}\right)\right]}\label{eq:Vesc}
\end{equation}
$V_{\rm max}$ being the maximum rotation velocity of the galaxy and $R_{\rm vir}$ its virial radius.
The virial radius of the galaxies can be define as $R_{\rm vir} \approx V_{\rm max}/10H(z)$ where $H(z)$ is the Hubble parameter at redshift $z$. 
In Table~\ref{table:results}, we compare the outflow velocity with the escape velocity for the wind-pair. 
This ratio $V_{\rm out}/V_{\rm esc}$ of 0.52 shows that 
the outflowing material is not able to reach the escape velocity and will thus likely to fall back onto the galaxy, assuming we are tracing the gas going out of the galaxy. 
One can ask whether we are already tracing the gas falling back onto the galaxy. 
If this is the case, we should see another opposite component (with respect to the systemic velocity) in the absorption profile corresponding to the outflowing gas. 

Table~\ref{table:results} also lists the estimated outflow rate. 
The errors on the ejected mass rate $\dot M_{\rm out}$ are dominated by the ones on the gas column density $N_{\rm HI}$ and the SFR. 

\begin{table*}[h]
\centering
\caption{Results for the galaxy \textbf{J215200+0625G2}.}
\label{table:results}
\begin{tabular}{lccccccccc}
\hline
Galaxy		&$b$ (kpc)	& log($N_{\rm H}(b)$)	& $V_{\rm max}$		& $V_{\rm out}$	&$\theta_{\rm max}$	&SFR		&$\dot M_{\rm out}$	&$\frac{V_{\rm out}}{V_{\rm esc}}$	&$\eta$	\\
(1)		&(2)		&(3)			&(4)			&(5)		&(6)			&(7)		&(8)			&(9)					&(10)	\\
\hline
\textbf{J215200+0625G2}	&34.0		&19.7$\pm$0.07		&$140.8 \pm 51$		&$150 \pm 10$	&20 $\pm$ 5.0		&4.6$\pm$0.4	&1.7 $^{+1.1}_{-0.8}$	&0.52					&0.75	\\
		&		&			&			&		&			&		&1.1 $^{+0.9}_{-0.6}$	&					&0.49	\\
\hline
\end{tabular}\\
{
(1) Galaxy name;
(2) Impact parameter (kpc);
(3) Gas column density at the impact parameter (cm$^{-2}$);
(4) Maximum rotational velocity of the galaxy (\kms ); 
(5) Wind velocity (\kms );
(6) Cone opening angle (degrees)
(7) Star Formation Rate (\mpy);
(8) Ejected mass rate for one cone (\mpy);
(9) Ejection velocity divided by escape velocity;
(10) Mass loading factor: ejected mass rate divided by star formation rate 
(for both cones).
Values in the second row ($\dot M_{\rm out}=1.1^{+0.9}_{-0.6}$~\mpy and $\eta=0.49$) correspond to the empty inner cone model.
\\
}
\end{table*}

From the outflow rate, we compute the mass loading factor $\eta$ by comparing it to the SFR ($\eta=\dot M_{\rm out} / \rm SFR$).
For our \textbf{SDSSJ215200+0625G2} pair, we used the empty cone model to reproduce the absorption profile with an inner cone opening angle $\theta_{\rm in}$ of 7$^\circ$. 
To be consistent with the other cases, we give two solutions for this galaxy$-$quasar pair: one with the filled cone and one with the inner cone subtracted.

\begin{figure*}[]
  \centering
  \includegraphics[width=14cm]{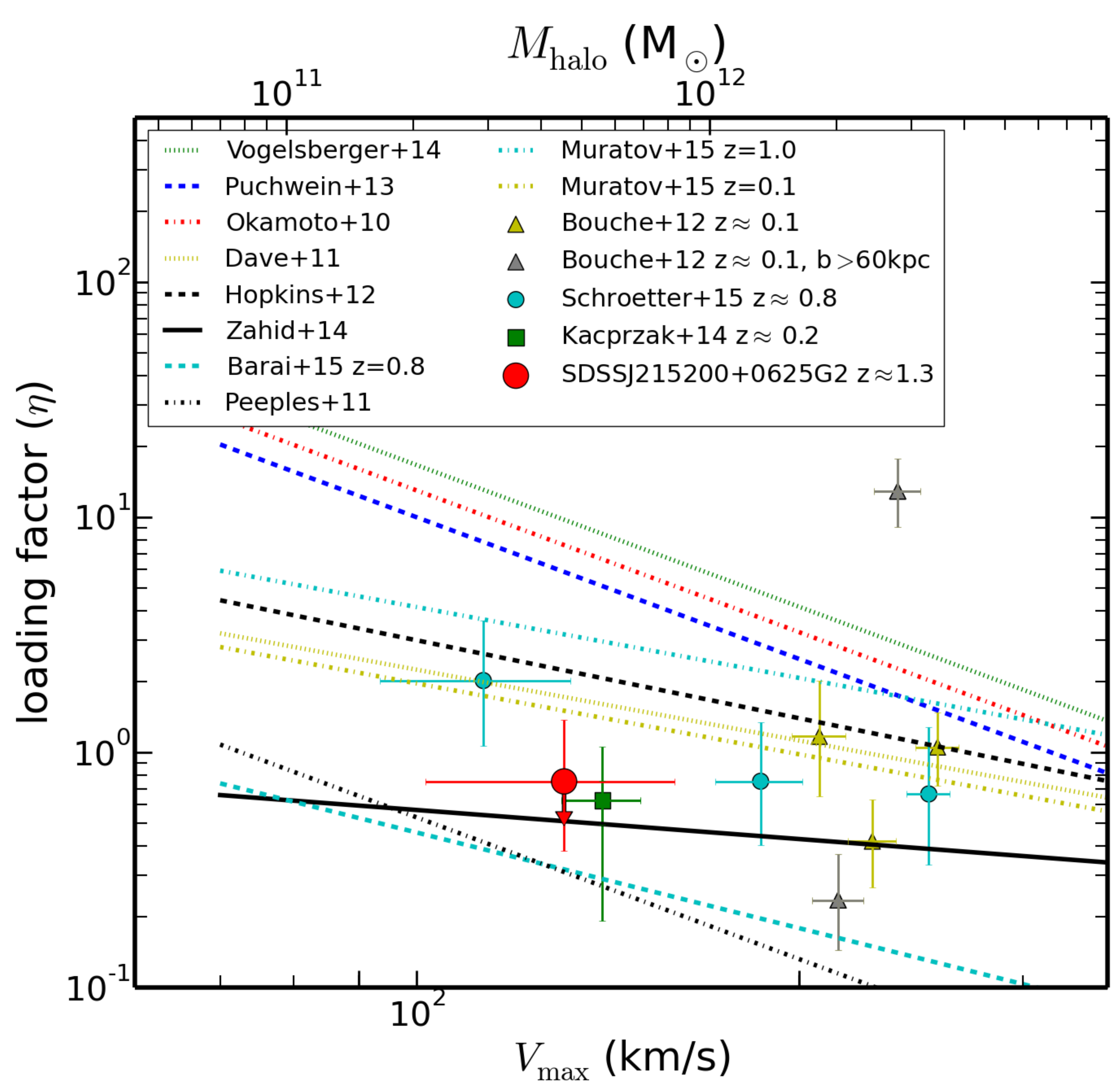}
  \caption{Comparison of mass loading factors assumed by theoretical/empirical models (curves) with values derived from background quasar observations 
  (dots and triangles) as a function of the maximum rotational velocity. 
 The result from this work is represented by the red circle.
 The red arrow represents the loading factor of the \textbf{SDSSJ215200+0625G2} galaxy with the subtracted mass from the inner cone model.
 The cyan circles show the results for galaxies at $z\approx0.8$ from \citet{schroetter_15}. 
 The green square shows the mass loading factor for a $z\approx0.2$ galaxy \citep{kacprzak_14}. 
 The triangles show the results for $z\approx0.2$ galaxies from \citet{bouche_12}.
 The gray triangles show the galaxies with quasars located at $>$60kpc where the mass loading factor is less reliable due to the large travel time 
 needed for the outflow to cross the quasar LOS (several 100 Myr) compared to the short time scale of the \Ha\ derived SFR ($\sim 10$Myr).
 The upper halo mass axis is scaled on $V_{\rm max}$ at redshift 0.8 from \citet{mo_02}.
  }
  \label{fig:eta_vs_vmax}
\end{figure*}

Figure~\ref{fig:eta_vs_vmax} shows the loading factor $\eta$ as a function of halo mass and maximum rotational velocity \Vmax\ for this work and previous similar studies \citep{bouche_12, kacprzak_14, schroetter_15}.
The derived loading factor for galaxy \textbf{SDSSJ215200+0625G2} follows the same trend as the others. 
The red arrow shows the loading factor for the subtracted mass from the low-density inner cone.

MUSE allows us to probe galaxies with an impact parameter larger than before with an IFU. 
But, in Figure~\ref{fig:eta_vs_vmax}, we caution the reader that loading factor for galaxies with impact parameters larger than 60 kpc are less
reliable because of the time needed for the gas to travel from the galaxy to the quasar LOS ($\sim 400$ Myr at $\Vout\approx 150 \kms$ with $b=60$ kpc). 
A major limitation for the comparison between data and models in Figure~\ref{fig:eta_vs_vmax},
is that $\eta$ in simulations are usually measured on a scale of a few kpc away from the galaxy, which is one order of magnitude lower than most of the observations (tens of kpc).

\section{Summary}
\label{conclusion}
We present results on 2 GTO VLT/MUSE fields in which we searched for galaxy-quasar pairs. 
These fields were selected from the SDSS database where we searched for multiple \MgII\ absorbers, with $z\approx 0.8 - 1.4$ and $W_r^{\lambda 2796}>0.5$\AA, 
in the quasar spectra. 
Out of 8 \MgII\ absorptions in the quasar spectra of these two fields, we detect 6 redshift-corresponding SFGs. 
For these 2 fields (J213748+1112 and J215200+0625) we also have high resolution spectra of the quasars from the VLT/UVES instrument. 
In each of these two fields, we detected more than 40 emitters in the $1\arcmin \times 1\arcmin$ MUSE field of view (see the Appendix). 
We focused on galaxies at MgII absorptions redshifts in the quasar spectra and for which the associated quasar LOS is aligned with their minor axis ($\alpha>55^\circ$) 
and is thus likely to probe outflowing materials (wind$-$pairs).
Among the 6 detected SFGs, one is likely to be a wind-pair due to its orientation with respect to its relative quasar. 

In summary, thanks to our new GTO VLT/MUSE and VLT/UVES data, 
MUSE allows us to detect galaxies far away from their associated quasar ($\sim 100$~kpc) as compare to previous similar works \citep[i.e. ][]{bouche_12, kacprzak_14, schroetter_15}.
For the wind-pair \textbf{SDSSJ215200+0625G2}, we found that the outflow velocity \Vout\ is $\approx 150$~\kms.
The outflowing gas is likely to stay inside the gravitational well of the galaxy and the loading factor is $\eta\approx0.7$.
We showed a gap in velocities in the absorption profile which led to a low-density inner cone modeling.
At this point, we have outflowing constraints for one galaxy but we showed that MUSE is able to provide very good data and will play a fundamental role in this field.

MUSE allowed us to probe multiple galactic wind cases at the same time and enhance the number of cases with only two quasar fields. 
We also have a case of low-density inner cone which opens discussions on geometrical properties of outflowing materials.
The MEGAFLOW sample is currently growing and successful in detecting galaxies in each quasar field ($\approx$84$\%$ detection). 
Future work will be done with a lot more observation with MUSE$+$UVES, and in a short time, the MEGAFLOW sample should be large enough to produce statistical results on outflow properties. 
\acknowledgments
\textit{Acknowledgments.} NB acknowledges support from a Career Integration Grant (CIG) (PCIG11-GA-2012-321702) within the 7th European Community Framework Program.
This work has been carried out thanks to the support of the ANR FOGHAR (ANR-13-BS05-0010-02), the OCEVU Labex (ANR-11-LABX-0060) and the A*MIDEX project (ANR-11-IDEX-0001-02) 
funded by the ``Investissements d'Avenir" French government program managed by the ANR.
This work received financial support from the European Research Council under
the European Union’s Seventh Framework Programme (FP7$/$2007-2013) $/$ ERC Grant agreement 278594-GasAroundGalaxies.

\pagebreak
\appendix
\setcounter{figure}{0} \renewcommand{\thefigure}{A.\arabic{figure}} 

\section{MUSE fields emitters detection}

For completeness we looked for these emitters by visual inspection and found $42$ galaxies with emission lines in each of these two fields 
(see Table~\ref{table:catalog_j2137} for SDSSJ213748$+$0012 and Table~\ref{table:catalog_j2152} for SDSSJ215200$+$0625).

\begin{table*}[]
\centering
\caption{MUSE Sources in the SDSSJ213748$+$0012 field with redshifts. 
Within these 42 emitters, $36$ have identified emission lines.}
\label{table:catalog_j2137}
\begin{tabular}{ccccc}
ID & R.A. & Dec. & redshift & lines \\
obj001 & 21:37:48.303 & +00:12:21.69 & 0.132 & OIII, H$_\beta$, H$_\alpha$, NII \\
obj002 & 21:37:48.757 & +00:12:19.29 & 0.156 & OIII, H$_\beta$, H$_\alpha$, NII \\
obj003 & 21:37:50.157 & +00:12:52.89 & 0.315 & H$_\beta$, OIII \\
obj004 & 21:37:48.370 & +00:12:23.89 & 0.325 & OII, OIII, H$_\beta$, H$_\alpha$, NII \\
obj005 & 21:37:48.370 & +00:12:24.09 & 0.325 & OII, OIII, H$_\beta$, H$_\alpha$ \\
obj006 & 21:37:48.930 & +00:12:38.69 & 0.409 & OII, OIII, H$_\alpha$, NII \\
obj007 & 21:37:49.223 & +00:12:20.09 & 0.410 & OII, OIII, H$_\beta$ \\
obj008 & 21:37:49.810 & +00:12:15.69 & 0.442 & OII \\
obj009 & 21:37:48.477 & +00:12:30.09 & 0.543 & OII, OIII, H$_\beta$ \\
obj010 & 21:37:48.450 & +00:12:29.49 & 0.543 & OII, OIII, H$_\beta$ \\
obj011 & 21:37:50.450 & +00:12:02.89 & 0.580 & OIII, H$_\beta$ \\
obj012 & 21:37:48.983 & +00:12:55.09 & 0.616 & OII, OIII, H$_\beta$ \\
obj013 & 21:37:49.343 & +00:12:52.09 & 0.684 & OII, OIII, H$_\beta$ \\
obj014 & 21:37:47.743 & +00:12:46.69 & 0.711 & OII \\
obj015 & 21:37:49.530 & +00:12:14.69 & 0.766 & OII \\
obj016 & 21:37:48.317 & +00:12:15.69 & 0.767 & OII \\
obj017 & 21:37:49.463 & +00:12:16.49 & 0.767 & OII, OIII \\
obj018 & 21:37:49.023 & +00:12:27.29 & 0.806 & OII, OIII, H$_\beta$ \\
obj019 & 21:37:48.823 & +00:12:27.49 & 0.806 & OII, OIII \\
obj020 & 21:37:50.157 & +00:12:30.89 & 0.806 & OII, OIII \\
obj021 & 21:37:49.490 & +00:12:33.69 & $\cdots$ & 8281.3 \\
obj022 & 21:37:50.103 & +00:12:53.29 & $\cdots$ & 6823. \\
obj023 & 21:37:49.117 & +00:12:11.89 & $\cdots$ & 6897. \\
obj024 & 21:37:47.663 & +00:12:12.69 & 0.900 & OII\\
obj025 & 21:37:48.930 & +00:12:09.49 & 0.902 & OII?\\
obj026 & 21:37:48.517 & +00:12:05.69 & $\cdots$ & 7079.69\\
obj027 & 21:37:48.063 & +00:12:33.69 & $\cdots$ & 7376.81\\
obj028 & 21:37:48.437 & +00:12:46.29 & 1.010 & OII \\
obj029 & 21:37:48.837 & +00:12:42.69 & 1.010 & OII \\
obj030 & 21:37:48.970 & +00:12:09.49 & 1.045 & OII \\
obj031 & 21:37:49.970 & +00:12:09.09 & 1.044 & OII \\
obj032 & 21:37:49.970 & +00:12:15.29 & 1.122 & OII \\
obj033 & 21:37:48.903 & +00:12:17.69 & 1.188 & OII \\
obj034 & 21:37:46.837 & +00:12:02.89 & 1.212 & OII \\
obj035 & 21:37:47.970 & +00:12:29.09 & 1.213 & OII \\
obj036 & 21:37:46.943 & +00:12:08.89 & 1.214 & OII \\
obj037 & 21:37:47.850 & +00:12:33.49 & 1.214 & OII \\
obj038 & 21:37:50.410 & +00:12:20.09 & 1.257 & OII \\
obj039 & 21:37:48.370 & +00:12:04.69 & 1.300 & OII \\
obj040 & 21:37:47.717 & +00:12:46.89 & $\cdots$ & 8569.12 \\
obj041 & 21:37:48.730 & +00:12:15.29 & 5.941 & 8434.53 Ly$\alpha$? \\
obj042 & 21:37:48.823 & +00:12:27.49 & 6.442 & 9043.03 Ly$\alpha$? \\
\end{tabular}

\end{table*}

\begin{figure}[]
  \centering
  \includegraphics[width=11cm]{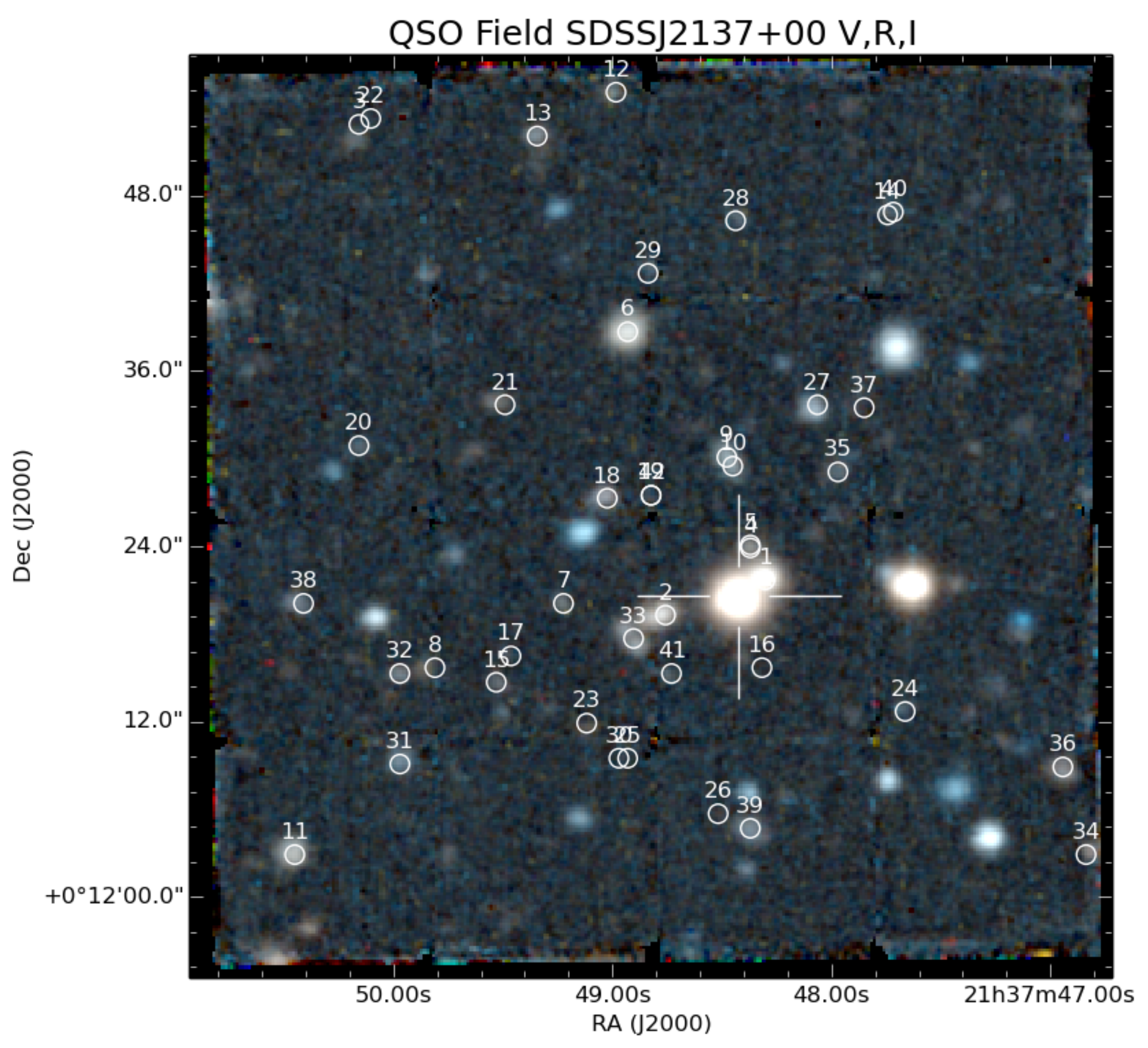}
  \caption{RGB image of the J213748+0012 field with identifications of emission detected galaxies.
  The white cross points the quasar location.
  Circles represent emission detected galaxies corresponding to Table~\ref{table:catalog_j2137}. 
  Not all the galaxy-like spots are circled on the image. 
  These spots are either stars or galaxies with a continuum but without obvious emission line.
  }
  \label{fig:J213748_field}
\end{figure}

\begin{table*}[]
\centering
\caption{MUSE Sources in the SDSSJ215200$+$0625 field with redshifts. 
We found 40 galaxies out of 41 having identified emission lines}
\label{table:catalog_j2152}
\begin{tabular}{ccccc}
ID & R.A. & Dec. & redshift & lines \\
obj001 & 21:52:02.018 & +06:25:47.66 & 0.433 & OII, OIII, H$_\beta$ \\
obj002 & 21:52:02.246 & +06:25:25.06 & 0.439 & OII \\
obj003 & 21:51:58.905 & +06:25:20.26 & 0.452 & OII, OIII, H$_\beta$ \\
obj004 & 21:52:02.085 & +06:25:13.26 & 0.489 & OII, OIII, H$_\beta$ \\
obj005 & 21:51:58.409 & +06:24:54.86 & 0.517 & OII \\
obj006 & 21:51:59.429 & +06:25:43.06 & 0.554 & OII, OIII \\
obj007 & 21:52:02.273 & +06:24:56.06 & 0.597 & OII, OIII, H$_\beta$ \\
obj008 & 21:52:00.770 & +06:25:17.26 & 3.931? & 5992.37 Ly$\alpha$?\\
obj009 & 21:51:59.200 & +06:24:54.86 & 4.196? & 6314.05 Ly$\alpha$?\\
obj010 & 21:51:58.878 & +06:25:01.46 & 0.742 & OII, OIII, H$_\beta$\\
obj011 & 21:51:59.912 & +06:25:15.66 & 0.748 & OII, H$_\beta$\\
obj012 & 21:52:02.139 & +06:25:31.26 & 0.770 & OII, OIII, H$_\beta$\\
obj013 & 21:51:59.375 & +06:25:40.26 & 0.786 & OII\\
obj014 & 21:52:00.341 & +06:25:22.46 & 0.332 & OII, OIII, H$_\alpha$ \\
obj015 & 21:52:01.092 & +06:25:16.26 & 0.824 & OII, OIII \\
obj016 & 21:52:00.636 & +06:25:37.66 & 0.289 & H$_\alpha$, NII \\
obj017 & 21:51:58.597 & +06:25:11.86 & 0.847 & OII? \\
obj018 & 21:51:59.818 & +06:25:29.66 & 0.873 & OII \\
obj019 & 21:52:00.126 & +06:25:13.06 & 0.879 & OII, OIII \\
obj020 & 21:52:00.234 & +06:24:50.86 & 0.438 & OII, OIII, H$_\beta$ \\
obj021 & 21:51:59.630 & +06:25:40.46 & 0.943 & OII \\
obj022 & 21:52:00.287 & +06:25:06.46 & 0.989 & OII \\
obj023 & 21:52:02.058 & +06:25:40.46 & 1.013 & OII \\
obj024 & 21:51:58.436 & +06:25:04.46 & 1.013 & OII \\
obj025 & 21:52:00.381 & +06:25:20.46 & 1.052 & OII \\
obj026 & 21:51:59.549 & +06:25:39.06 & 1.053 & OII \\
obj027 & 21:52:02.380 & +06:24:58.06 & 0.185 & OIII, H$_\beta$, H$_\alpha$, NII \\
obj028 & 21:51:58.583 & +06:25:34.26 & $\cdots$ & 8413.87 \\
obj029 & 21:52:00.904 & +06:24:50.26 & 1.302 & OII \\
obj030 & 21:52:00.019 & +06:25:13.26 & 1.318 & OII \\
obj031 & 21:51:59.952 & +06:25:15.46 & 1.318 & OII \\
obj032 & 21:51:58.355 & +06:25:03.06 & 1.349 & OII \\
obj033 & 21:51:58.489 & +06:24:59.06 & $\cdots$ & 8757.32 \\
obj034 & 21:52:02.354 & +06:25:15.46 & 1.362 & OII \\
obj035 & 21:51:58.355 & +06:25:23.66 & 1.403 & OII \\
obj036 & 21:52:00.435 & +06:25:13.46 & 1.430 & OII \\
obj037 & 21:52:00.623 & +06:25:15.86 & 1.430 & OII \\
obj038 & 21:52:01.629 & +06:25:24.06 & 1.431 & OII \\
obj039 & 21:52:00.972 & +06:25:33.06 & 1.433 & OII \\
obj040 & 21:52:00.703 & +06:25:43.06 & 1.435 & OII \\
obj041 & 21:52:00.180 & +06:25:41.26 & 1.432 & OII \\
\end{tabular}

\end{table*}

\begin{figure}[]
  \centering
  \includegraphics[width=11cm]{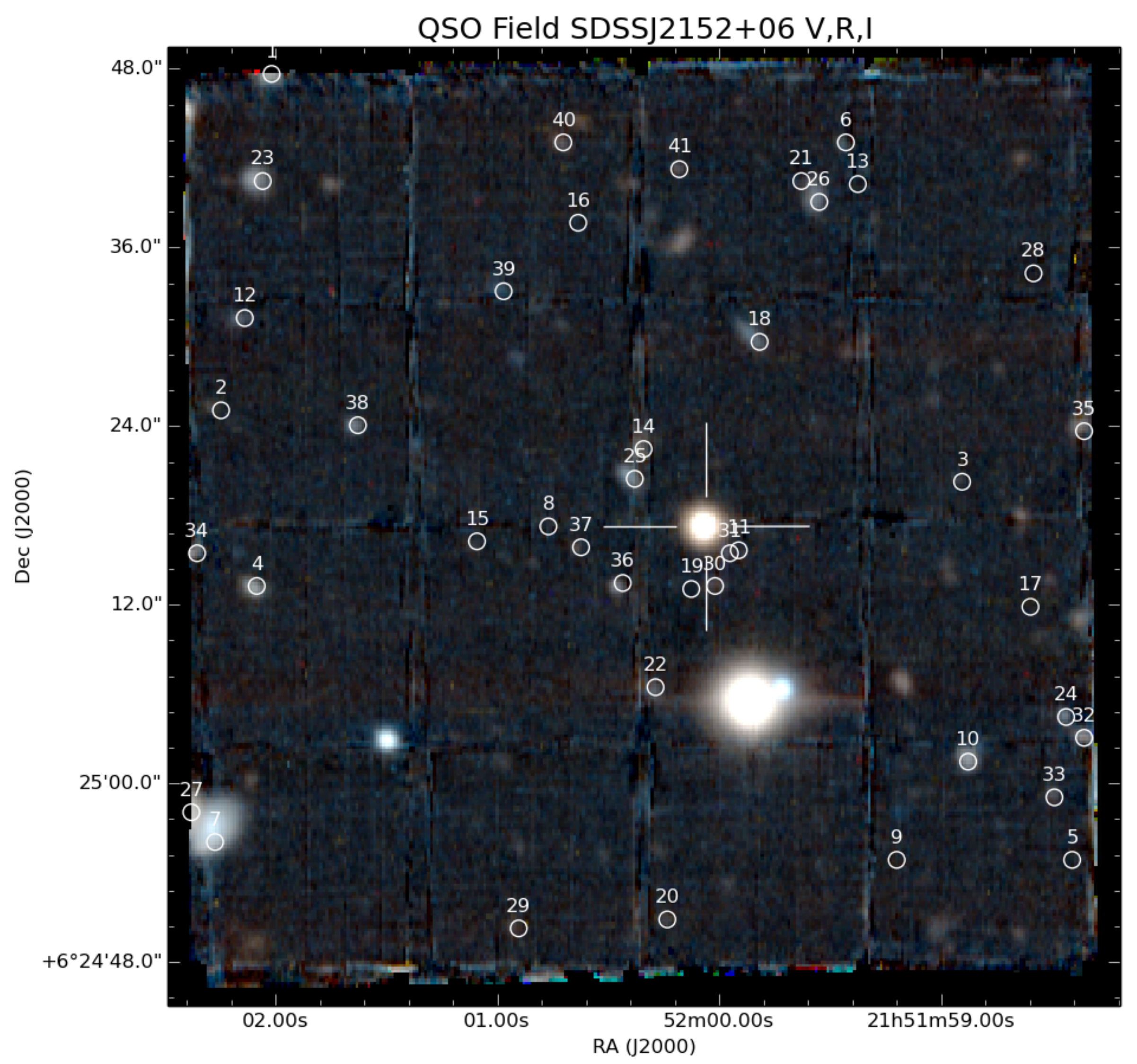}
  \caption{Same as Figure~\ref{fig:J213748_field} but for the J215200+0625 quasar field.
  Again, the white cross shows the quasar location and galaxies with emission lines are circled and listed in Table~\ref{table:catalog_j2152}.
  }
  \label{fig:J215200_field}
\end{figure}


\end{document}